\documentclass[letterpaper,english,american,prl,twocolumn,amsmath,amssymb,superscriptaddress]{revtex4-2}
\usepackage[T1]{fontenc}
\usepackage{xcolor}
\usepackage{babel}
\usepackage{prettyref}
\usepackage{bm}
\usepackage{amsmath}
\usepackage{amssymb}
\usepackage{graphicx}
\usepackage[unicode=true,
 bookmarks=true,bookmarksnumbered=false,bookmarksopen=false,
 breaklinks=false,pdfborder={0 0 1},backref=false,colorlinks=false]
 {hyperref}
\hypersetup{pdftitle={Large Deviation Approach to Random Recurrent Neuronal Networks: Parameter Inference and Fluctuation-Induced Transitions},
 pdfauthor={Alexander van Meegen, Tobias Kuehn, Moritz Helias},
 colorlinks=true,linkcolor=black,citecolor=black,urlcolor=black,filecolor=black}

\makeatletter

\pdfpageheight\paperheight
\pdfpagewidth\paperwidth

\usepackage[latin1]{inputenc}

\newrefformat{eq}{\hyperref[#1]{Eq.~(\ref{#1})}}
\newrefformat{cap}{\hyperref[#1]{Fig.~\ref{#1}}}
\newrefformat{fig}{\hyperref[#1]{Fig.~\ref{#1}}}
\newrefformat{tab}{\hyperref[#1]{Table ~\ref{#1}}}
\newrefformat{sec}{\hyperref[#1]{Section~\ref{#1}}}
\newrefformat{sub}{\hyperref[#1]{Section~\ref{#1}}}
\newrefformat{cha}{\hyperref[#1]{Chapter~\ref{#1}}}

\usepackage{MnSymbol}
\DeclareMathAlphabet\mathbb{U}{msb}{m}{n}

\makeatother

\begin{document}
\selectlanguage{english}%
\global\long\def\T{\!\mathsf{T}}%
\global\long\def\l{\langle}%
\global\long\def\r{\rangle}%
\global\long\def\ll{\langle\langle}%
\global\long\def\rr{\rangle\rangle}%
\global\long\def\tr{\mathrm{tr}}%
\global\long\def\lr#1{\left\langle #1\right\rangle }%
\global\long\def\vect#1{\bm{#1}}%
\global\long\def\matr#1{\bm{#1}}%
\global\long\def\uint#1{\int\mathrm{d}#1\,}%
\global\long\def\dint#1#2#3{\int_{#2}^{#3}\mathrm{d}#1\,}%
\global\long\def\pint#1{\int\mathcal{D}#1\,}%
\global\long\def\C{C}%
\global\long\def\Ct{\tilde{C}}%
\global\long\def\k{\ell}%
\global\long\def\evalat#1{\left.#1\right|}%
\global\long\def\xt{\tilde{x}}%
\global\long\def\erf{\mathrm{erf}}%
\global\long\def\clip{\mathrm{clip}}%

\title{Large Deviations Approach to Random Recurrent Neuronal Networks:\\
Parameter Inference and Fluctuation--Induced Transitions}
\author{Alexander van Meegen}
\affiliation{Institute of Neuroscience and Medicine (INM-6) and Institute for Advanced
Simulation (IAS-6) and JARA-Institute Brain Structure-Function Relationships
(INM-10), Jülich Research Centre, Jülich, Germany}
\affiliation{Institute of Zoology, University of Cologne, 50674 Cologne, Germany}
\author{Tobias Kühn}
\affiliation{Institute of Neuroscience and Medicine (INM-6) and Institute for Advanced
Simulation (IAS-6) and JARA-Institute Brain Structure-Function Relationships
(INM-10), Jülich Research Centre, Jülich, Germany}
\affiliation{Department of Physics, Faculty 1, RWTH Aachen University, Aachen,
Germany}
\affiliation{Laboratoire de Physique de l'ENS, Laboratoire MSC de l'Université
de Paris, CNRS, Paris, France}
\author{Moritz Helias}
\affiliation{Institute of Neuroscience and Medicine (INM-6) and Institute for Advanced
Simulation (IAS-6) and JARA-Institute Brain Structure-Function Relationships
(INM-10), Jülich Research Centre, Jülich, Germany}
\affiliation{Department of Physics, Faculty 1, RWTH Aachen University, Aachen,
Germany}
\date{\today}
\begin{abstract}

We here unify the field theoretical approach to neuronal networks
with large deviations theory. For a prototypical random recurrent
network model with continuous-valued units, we show that the effective
action is identical to the rate function and derive the latter using
field theory. This rate function takes the form of a Kullback-Leibler
divergence which enables data-driven inference of model parameters
and calculation of fluctuations beyond mean--field theory. Lastly,
we expose a regime with fluctuation--induced transitions between
mean--field solutions.

\end{abstract}
\maketitle

\paragraph*{Introduction.--}

Biological neuronal networks are systems with many degrees of freedom
and intriguing properties: their units are coupled in a directed,
non-symmetric manner, so that they typically operate outside thermodynamic
equilibrium \citep{Rabinovich06_1213,Sompolinsky88_2}. The primary
analytical method to study neuronal networks has been mean-field theory
\citep{Amari72_643,Sompolinsky88_259,Stern14_062710,Kadmon15_041030,Aljadeff15_088101,vanMeegen18_258302}.
Its field-theoretical basis has been exposed only recently \citep{Crisanti18_062120,Schuecker18_041029}.
However, to understand the parallel and distributed information processing
performed by neuronal networks, the study of the forward problem --
from the microscopic parameters of the model to its dynamics -- is
not sufficient. One additionally faces the inverse problem of determining
the parameters of the model given a desired dynamics and thus function.
Formally, one needs to link statistical physics with concepts from
information theory and statistical inference.

We here expose a tight relation between statistical field theory of
neuronal networks, large deviations theory, information theory, and
inference. To this end, we generalize the probabilistic view of large
deviations theory, which yields rigorous results for the leading order
behavior in the network size $N$ \citep{Ben-Arous95_455,Guionnet97_183},
to arbitrary single unit dynamics, transfer functions, and multiple
populations. We furthermore show that the central quantity of large
deviations theory, the rate function, is identical to the effective
action in statistical field theory. This link exposes a second relation:
Bayesian inference and prediction are naturally formulated within
this framework, spanning the arc to information processing. Concretely,
we develop a method for parameter inference from transient data for
single- and multi-population networks. Lastly, we overcome the inherent
limit of mean-field theory---its neglect of fluctuations. We develop
a theory for fluctuations of the order parameter when the intrinsic
timescale is large and discover a regime with fluctuation--induced
transitions between two coexisting mean--field solutions.

First, we introduce the model in its most general form. Then, we develop
the theory for a single population. Last, we generalize it to multiple
populations.

\paragraph{Model.-}

We consider block-structured random networks of $N=\sum_{\alpha}N_{\alpha}$
nonlinearly interacting units $x_{i}^{\alpha}(t)$ driven by an external
input $\xi_{i}^{\alpha}(t)$. The dynamics of the $i$-th unit in
the $\alpha$-th population is governed by the stochastic differential
equation

\begin{align}
\tau_{\alpha}\dot{x}_{i}^{\alpha}(t) & =-U_{\alpha}^{\prime}(x_{i}^{\alpha}(t))+\sum_{\beta}\sum_{j=1}^{N_{\beta}}J_{ij}^{\alpha\beta}\phi(x_{j}^{\beta}(t))+\xi_{i}^{\alpha}(t).\label{eq:network_ode}
\end{align}
In the absence of recurrent and external inputs, the units undergo
an overdamped motion with time constant $\tau_{\alpha}$ in a potential
$U_{\alpha}(x)$. The $J_{ij}^{\alpha\beta}$ are independent and
identically Gaussian-distributed random coupling weights with zero
mean and population-specific variance $\l(J_{ij}^{\alpha\beta})^{2}\r=g_{\alpha\beta}^{2}/N_{\beta}$
where the coupling strength $g_{\alpha\beta}$ controls the heterogeneity
of the weights. The time-varying external inputs $\xi_{i}^{\alpha}(t)$
are independent Gaussian white-noise processes with zero mean and
correlation functions $\lr{\xi_{i}^{\alpha}(t_{1})\xi_{j}^{\beta}(t_{2})}=2D_{\alpha}\delta_{ij}\delta_{\alpha\beta}\delta(t_{1}-t_{2})$.
The single-population model corresponds to the one studied in Ref.~\citep{Sompolinsky88_259}
if the external input vanishes, $D=0$, the potential is quadratic,
$U(x)=\frac{1}{2}x^{2}$, and the transfer function is sigmoidal,
$\phi(x)=\tanh(x)$; for $D=\frac{1}{2}$, $U(x)=-\log(A^{2}-x^{2})$,
and $\phi(x)=x$ it corresponds to the one in Ref.~\citep{Ben-Arous95_455},
which is inspired by the dynamical spin glass model of Ref.~\citep{Sompolinsky81}.

\paragraph*{Field theory.-}

The field-theoretical treatment of \prettyref{eq:network_ode} employs
the Martin--Siggia--Rose--de Dominicis--Janssen path integral
formalism \citep{Martin73,janssen1976_377,Chow15,Hertz16_033001}.
We denote the expectation over paths across different realizations
of the noise $\xi$ as (Appendix A1)
\begin{align*}
\lr{\cdot}_{\vect x|\matr J} & \equiv\lr{\lr{\cdot}_{\vect x|\matr J,\vect{\xi}}}_{\vect{\xi}}=\pint{\vect x}\pint{\vect{\tilde{x}}}\cdot\,e^{S_{0}(\vect x,\vect{\tilde{x}})-\vect{\tilde{x}}^{\T}\matr J\phi(\vect x)},
\end{align*}
where $\lr{\cdot}_{\vect x|\matr J,\vect{\xi}}$ integrates over the
unique solution of \prettyref{eq:network_ode} given one realization
$\vect{\xi}$ of the noise. Here, $S_{0}(\vect x,\vect{\tilde{x}})=\vect{\tilde{x}}^{\T}(\dot{\vect x}+U^{\prime}(\vect x))+D\vect{\tilde{x}}^{\T}\vect{\tilde{x}}$
is the action of the uncoupled neurons. We use the shorthand notation
$\vect a^{\T}\vect b=\sum_{i=1}^{N}\dint t0Ta_{i}(t)b_{i}(t)$.

For large $N$, the system becomes self-averaging, a property known
from many disordered systems with large numbers of degrees of freedom:
the collective behavior is stereotypical, independent of the realization
$J_{ij}$. A self--averaging observable has a sharply peaked distribution
over realizations of $\matr J$---the observable always attains the
same value, close to its average. This, however, only holds for observables
averaged over all units, reminiscent of the central limit theorem.
These are generally of the form $\sum_{i=1}^{N}\k(x_{i})$, where
$\ell$ is an arbitrary functional of a single unit's trajectory.
It is therefore convenient to introduce the scaled cumulant--generating
functional
\begin{align}
W_{N}(\ell) & :=\frac{1}{N}\ln\lr{\lr{e^{\sum_{i=1}^{N}\k(x_{i})}}_{\vect x|\matr J}}_{\matr J},\label{eq:def_WN}
\end{align}
where the prefactor $1/N$ makes sure that $W_{N}$ is an intensive
quantity, reminiscent of the bulk free energy \citep{Goldenfeld92}.
In fact, we will show that the $N$-dependence vanishes in the limit
$N\to\infty$ because the system decouples.

Performing the average over $\matr J$, i.e.~evaluating $\l e^{-\vect{\tilde{x}}^{\T}\matr J\phi(\vect x)}\r_{\matr J}$,
and introducing the auxiliary field
\begin{align}
C(t_{1},t_{2}) & :=\frac{1}{N}\sum_{i=1}^{N}\phi(x_{i}(t_{1}))\,\phi(x_{i}(t_{2}))\label{eq:def_aux}
\end{align}
 as well as the conjugate field $\tilde{C}$, we can write $W_{N}$
as (Appendix A1)

\begin{align}
W_{N}(\ell) & =\frac{1}{N}\ln\,\pint C\pint{\tilde{C}}e^{-N\,C^{\T}\tilde{C}+N\,\Omega_{\ell}(C,\tilde{C})},\label{eq:W_N}\\
\Omega_{\ell}(C,\tilde{C}) & :=\ln\,\pint x\pint{\xt}e^{S_{0}(x,\xt)+\frac{g^{2}}{2}\xt^{\T}C\xt+\phi^{\T}\tilde{C}\phi+\k(x)}.\nonumber 
\end{align}
The effective action is defined as the Legendre transform of $W_{N}(\ell)$,
\begin{align}
\Gamma_{N}(\mu) & :=\pint x\mu(x)\,\ell_{\mu}(x)-W_{N}(\ell_{\mu}),\label{eq:Gamma_N}
\end{align}
where $\ell_{\mu}$ is determined implicitly by the condition $\mu=W_{N}^{\prime}(\ell_{\mu})$
and the derivative $W_{N}^{\prime}(\ell)$ has to be understood as
a generalized derivative, the coefficient of the linearization akin
to a Fréchet derivative \citep{Berger77}.

Note that $W_{N}$ and $\Gamma_{N}$ are, respectively, generalizations
of a cumulant--generating functional and of the effective action
\citep{ZinnJustin96} because both map a functional ($\ell$ or $\mu$)
to the reals. For the choice $\ell(x)=j^{\T}x$, where $j(t)$ is
an arbitrary function, we recover the usual cumulant--generating
functional of the single unit's trajectory (Appendix A4) and the corresponding
effective action.

\paragraph*{Rate function.-}

Any network--averaged observable, for which we may expect self-averaging
to hold, can likewise be obtained from the empirical measure
\begin{align}
\mu(y) & :=\frac{1}{N}\sum_{i=1}^{N}\delta(x_{i}-y),\label{eq:emp_meas}
\end{align}
since $\frac{1}{N}\sum_{i=1}^{N}\ell(x_{i})=\pint y\mu(y)\ell(y)$.
Of particular interest is the leading--order exponential behavior
of the distribution of empirical measures $P(\mu)=\l\l P(\mu\,|\,\vect x)\r_{\vect x|\matr J}\r_{\matr J}$
across realizations of $\matr J$ and $\vect{\xi}$. This behavior
in the large $N$ limit is described by what is known as the rate
function
\begin{align}
H(\mu) & :=-\lim_{N\to\infty}\frac{1}{N}\ln P(\mu)\label{eq:rate_function}
\end{align}
in large deviations theory \citep[see e.g.][]{Mezard_Montanari09};
$H(\mu)$ captures the leading exponential probability $P(\mu)\stackrel{N\gg1}{\simeq}e^{-N\,H(\mu)}$.
For large $N$, the probability of an empirical measure that does
not correspond to the minimum $H^{\prime}(\bar{\mu})=0$ is thus exponentially
suppressed. Put differently, the system is self--averaging and the
statistics of any network--averaged observable can be obtained using
$\bar{\mu}$.

Similar as in field theory, it is convenient to introduce the scaled
cumulant--generating functional of the empirical measure. Because
$\frac{1}{N}\sum_{i=1}^{N}\ell(x_{i})=\pint y\mu(y)\ell(y)$ holds
for an arbitrary functional $\ell(x_{i})$ of the single unit's trajectory
$x_{i}$, \prettyref{eq:def_WN} has the form of the scaled cumulant--generating
functional for $\mu$ at finite $N$.

Using a saddle-point approximation for the integrals over $C$ and
$\tilde{C}$ in \prettyref{eq:W_N} (Appendix A1), we get
\begin{align}
W_{\infty}(\ell) & =-\C_{\k}^{\T}\Ct_{\k}+\Omega_{\k}(\C_{\k},\Ct_{\k}).\label{eq:scgf_spa}
\end{align}
Both $\C_{\k}$ and $\Ct_{\k}$ are determined self-consistently by
the saddle-point equations $\C_{\k}=\evalat{\partial_{\Ct}\Omega_{\k}(\C,\Ct)}_{\C_{\k},\Ct_{\k}}$
and $\Ct_{\k}=\evalat{\partial_{\C}\Omega_{\k}(\C,\Ct)}_{\C_{\k},\Ct_{\k}}$
where $\partial_{C}$ denotes a partial functional derivative.

From the scaled cumulant--generating functional, \prettyref{eq:scgf_spa},
we obtain the rate function via a Legendre transformation \citep{Touchette09}:
$H(\mu)=\pint x\mu(x)\k_{\mu}(x)-W_{\infty}(\ell)$ with $\k_{\mu}$
implicitly defined by $\mu=W_{\infty}^{\prime}(\k_{\mu})$. Note that
$H(\mu)$ is still convex even if $\mu$ itself is multimodal. Comparing
with \prettyref{eq:Gamma_N}, we observe that the rate function is
equivalent to the effective action: $H(\mu)=\lim_{N\to\infty}\Gamma_{N}(\mu)$.
The equation $\mu=W_{\infty}^{\prime}(\k_{\mu})$ can be solved for
$\k_{\mu}$ to obtain a closed expression for the rate function viz.~effective
action (Appendix A2), one main result of our work,
\begin{align}
H(\mu) & =\pint x\mu(x)\ln\frac{\mu(x)}{\lr{\delta(\dot{x}+U^{\prime}(x)-\eta)}_{\eta}},\label{eq:ratefct_emp_meas}
\end{align}
where $\eta$ is a zero--mean Gaussian process with a correlation
function that is determined by $\mu(x)$, 
\begin{align}
C_{\eta}(t_{1},t_{2})= & 2D\,\delta(t_{1}-t_{2})\nonumber \\
 & +g^{2}\negthinspace\pint x\mu(x)\,\phi(x(t_{1}))\phi(x(t_{2})).\label{eq:auto_corr_eta_empirical}
\end{align}
For $D=\frac{1}{2}$, $U(x)=-\log(A^{2}-x^{2})$, and $\phi(x)=x$,
\prettyref{eq:ratefct_emp_meas} can be shown to be a equivalent to
the mathematically rigorous result obtained in the seminal work by
Ben Arous and Guionnet (Appendix A3).

The rate function \prettyref{eq:ratefct_emp_meas} takes the form
of a Kullback-Leibler divergence. Thus, it possesses a minimum at
\begin{align}
\bar{\mu}(x) & =\lr{\delta(\dot{x}+U^{\prime}(x)-\eta)}_{\eta}.\label{eq:selfconsistent_mu}
\end{align}
This most likely measure corresponds to the well-known self-consistent
stochastic dynamics that is obtained in field theory \citep{Sompolinsky88_259,Crisanti18_062120,Schuecker18_041029,Helias20_970}.
Note that the correlation function of the effective stochastic input
$\eta$ at the minimum depends self-consistently on $\bar{\mu}(x)$
through \prettyref{eq:auto_corr_eta_empirical}. However, the rate
function $H(\mu)$ contains more information. It quantifies the suppression
of departures $\mu-\bar{\mu}$ from the most likely measure and therefore
allows the assessment of fluctuations that are beyond the scope of
the classical mean-field result.

\paragraph{Parameter Inference.--}

\begin{figure}
\includegraphics{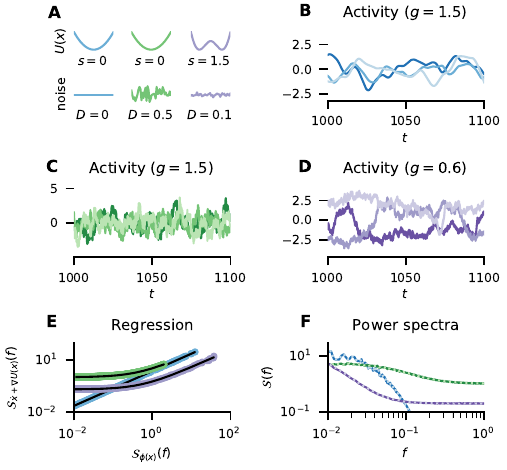}

\caption{Maximum likelihood parameter estimation for $\phi(x)=\protect\erf(\sqrt{\pi}x/2)$,
potential $U(x)=\frac{1}{2}x^{2}+s\ln\cosh x$, and external noise
$D$. \textbf{A} Color--coded sketch of potential and noise. \textbf{B}-\textbf{D}
Activity of three randomly chosen units for coupling strengths $g$
indicated in title. \textbf{E} Parameter estimation via non-negative
least squares regression (black lines) based on \prettyref{eq:inference_condition_stat}.
\textbf{F} Power spectra on the left-- (dark, solid curves) and right--hand--sides
(light, dotted curves) of \prettyref{eq:inference_condition_stat}
for the inferred parameters. Further parameters: $N=10,000$, temporal
discretization $dt=10^{-2}$, simulation time $T=1,000$, time-span
discarded to reach steady state $T_{0}=100$. \label{fig:inference}}
\end{figure}

The rate function opens the way to address the inverse problem: given
the network--averaged activity statistics, encoded in the corresponding
empirical measure $\mu$, what are the statistics of the connectivity
and the external input, i.e.~$g$ and $D$?

We determine the parameters using maximum likelihood estimation. Using
\prettyref{eq:rate_function} and \prettyref{eq:ratefct_emp_meas},
the likelihood of the parameters is given by
\begin{align*}
\ln P(\mu\,|\,g,D) & \simeq-NH(\mu\,|\,g,D),
\end{align*}
where $\simeq$ denotes equality in the limit $N\to\infty$ and we
made the dependence on $g$ and $D$ explicit. The maximum likelihood
estimate of the parameters $g$ and $D$ corresponds to the minimum
of the Kullback--Leibler divergence $H$, \prettyref{eq:ratefct_emp_meas},
on the right hand side. Evaluating the derivative of $H(\mu\,|\,g,D)$
yields (Appendix B1)
\begin{align*}
\partial_{a}\ln P(\mu\,|\,g,D) & \simeq-\frac{N}{2}\tr\left((C_{0}-C_{\eta})\frac{\partial C_{\eta}^{-1}}{\partial a}\right),
\end{align*}
where we abbreviated $a\in\{g,D\}$ and defined $C_{0}(t_{1},t_{2})\equiv\pint x\mu(x)\,\big(\dot{x}(t_{1})+U^{\prime}(x(t_{1}))\big)\,\big(\dot{x}(t_{2})+U^{\prime}(x(t_{2}))\big)$.
The derivative vanishes for $C_{0}=C_{\eta}$. Assuming stationarity,
in Fourier domain this condition reads
\begin{align}
\mathcal{S}_{\dot{x}+U^{\prime}(x)}(f) & =2D+g^{2}\mathcal{S}_{\phi(x)}(f),\label{eq:inference_condition_stat}
\end{align}
where $\mathcal{S}_{X}(f)$ denotes the network--averaged power spectrum
of the observable $X$. Using non--negative least squares \citep{Lawson95},
\prettyref{eq:inference_condition_stat} allows a straightforward
inference of $g$ and $D$ (\prettyref{fig:inference}). To determine
the transfer function $\phi$ and the potential $U$, one can use
model comparison techniques (Appendix B2). Using the inferred parameters,
we can also predict the future activity of a unit from the knowledge
of its recent past (Appendix B3).

\paragraph{Fluctuations.--}

\begin{figure}
\includegraphics{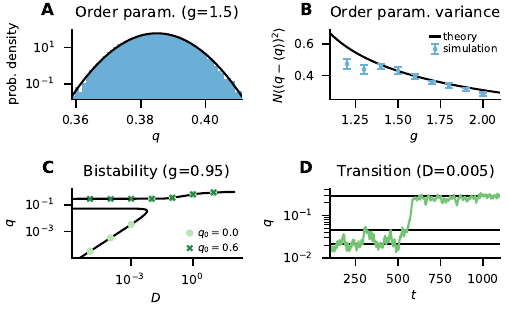}\caption{Order parameter fluctuations for $\phi(x)=\protect\erf(\sqrt{\pi}x/2)$
(\textbf{A},\textbf{B}) and meta-stability for $\phi(x)=\protect\clip(\tan(x),-1,1)$
(\textbf{C},\textbf{D}). \textbf{A} Temporal order parameter statistics
across ten simulations (bars) and theory (solid curve) from \prettyref{eq:Deltaq}.
\textbf{B} Order parameter variance for $10$ realizations of the
connectivity with standard error of the mean (symbols) and theory
(solid curve) from \prettyref{eq:Deltaq}. \textbf{C} Mean order parameter
for different initial values $q_{0}$ from simulations (symbols) and
self--consistent theory (solid curves). \textbf{D} Fluctuation induced
bistability of the order parameter for $N=750$, $g=0.95$. $T=5,000$
in \textbf{A},\textbf{D}; $U(x)=\frac{1}{2}x^{2}$; further parameters
as in \prettyref{fig:inference}.\label{fig:fluctuations}}
\end{figure}
The rate function allows us to go beyond mean--field theory and examine
fluctuations of the order parameter. Here, we use the network-averaged
variance $q(t)=C(t,t)$ from \prettyref{eq:def_aux} as an order parameter
and restrict the discussion to the case $U(x)=\frac{1}{2}x^{2}.$

\prettyref{fig:fluctuations}\textbf{A} shows the distribution of
$q(t)$ across time and across realizations of the connectivity.
The fluctuations across realizations of the connectivity can be computed
from the curvature of the rate function $I(C)$ that is obtained
from \eqref{eq:ratefct_emp_meas} by the contraction principle (Appendix
C1). In a stationary state and considering only the fluctuations across
realizations of the connectivity, for slow recurrent dynamics $\tau_{c}\gg1$
we obtain the approximation for the fluctuations of $q$
\begin{align}
\langle(q-\langle q\rangle_{\matr J})^{2}\rangle_{\matr J} & =\frac{\l(\phi\phi-\l\phi\phi\r_{0})^{2}\r_{0}}{N\left(1-g^{2}\left(\l\phi^{\prime\prime}\phi\r_{0}+\l\phi^{\prime}\phi^{\prime}\r_{0}\right)\right)^{2}}.\label{eq:Deltaq}
\end{align}
Here, $\l fg\r_{0}\equiv\l f(x(t))g(x(t))\r_{0}$ denotes an expectation
w.r.t.~the self--consistent measure \eqref{eq:selfconsistent_mu}.
For vanishing noise, $D=0$, and $g>1$, the dynamics are slow and
the theory matches the empirical fluctuations very well (\prettyref{fig:fluctuations}\textbf{A},\textbf{B}).
Deviations in \prettyref{fig:fluctuations}\textbf{B} are caused by
two effects: For $g\searrow1$, periodic solutions appear as a finite-size
effect; for growing $g$, the timescale $\tau_{c}$ decreases, eventually
violating the assumption $\tau_{c}\gg1$ entering \prettyref{eq:Deltaq}.
Rate functions like $I(C)$ in general also allow one to estimate
the tail probability $\mathbb{P}(q>\theta)\approx\exp(-NI(\theta))$,
which here shows a quadratic decline for large departures (\prettyref{fig:fluctuations}\textbf{A}).

When the denominator in \prettyref{eq:Deltaq} vanishes, fluctuations
grow large, indicative of a continuous phase transition. For $\phi^{\prime\prime\prime}(0)<0$
the denominator vanishes for $g\ge1$ (\prettyref{fig:fluctuations}\textbf{B}),
in line with the established theory, the breakdown of linear stability
of the fixed point $x=0$ \citep{Sompolinsky88_259}. For $\phi^{\prime\prime\prime}(0)>0$,
however, \prettyref{eq:Deltaq} predicts qualitatively different behavior:
the denominator vanishes at $g<1$, in the linearly stable regime.
In fact, we find that this regime features the coexistence of two
stable mean--field solutions (\prettyref{fig:fluctuations}\textbf{C},
Appendix C2) and fluctuation-driven first order transitions between
them (\prettyref{fig:fluctuations}\textbf{D}). The solution with
larger $q$ corresponds to self--sustained activity; the solution
with smaller $q$ corresponds to the fixed point $x=0$ and is stable
(Appendix C2), in contrast to the case of a threshold-power-law transfer
function \citep{Kadmon15_041030}.

\paragraph{Multiple Populations.--}

\begin{figure}
\includegraphics{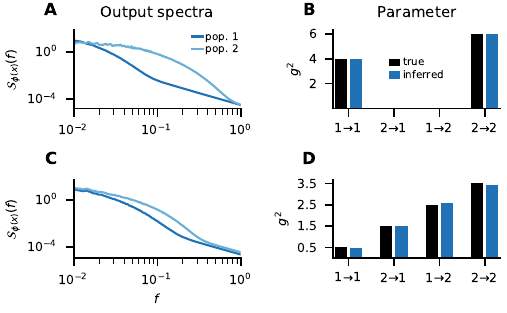}

\caption{Maximum likelihood parameter estimation for two populations with different
time constants $\tau_{1}=5$, $\tau_{2}=1$. \textbf{A} Output power
spectra $\mathcal{S}_{\phi(x)}^{\alpha}(f)$ of two unconnected populations
$g_{12}^{2}=g_{21}^{2}=0$ with $g_{11}^{2}=4$ and $g_{22}^{2}=6$.
\textbf{B} Estimated (blue) and true (black) parameters corresponding
to \textbf{A}. \textbf{C} Output power spectra of two connected populations
with $g_{11}^{2}=0.5$, $g_{12}^{2}=1.5$, $g_{21}^{2}=2.5$, and
$g_{22}^{2}=3.5$. \textbf{D} Estimated (blue) and true (black) parameters
corresponding to \textbf{C}. Further parameters: $N_{1}=N_{2}=5,000$,
$\phi(x)=\protect\erf(\sqrt{\pi}x/2)$, $U(x)=\frac{1}{2}x^{2}$,
and $D=0$; simulation parameters as in \prettyref{fig:inference}.\label{fig:inference_multipop}}
\end{figure}
For multiple populations, any population-averaged observable can be
obtained from the empirical measure $\mu^{\alpha}(y)=\frac{1}{N_{\alpha}}\sum_{i=1}^{N_{\alpha}}\delta(x_{i}^{\alpha}-y)$.
The joint distribution of all population-specific empirical measures
$\{\mu^{\circ}\}$ is determined by the rate function (Appendix D)
\begin{equation}
H(\{\mu^{\circ}\})=\sum_{\alpha}\gamma_{\alpha}\pint x\mu^{\alpha}(x)\,\ln\frac{\mu^{\alpha}(x)}{\lr{\delta(\tau_{\alpha}\dot{x}+U_{\alpha}^{\prime}(x)-\eta_{\alpha})}_{\eta_{\alpha}}}\,,\label{eq:rate_fct_multipop}
\end{equation}
where $\gamma_{\alpha}=N_{\alpha}/N$ and $\eta_{\alpha}$ is a zero-mean
Gaussian process with
\begin{align}
C_{\eta}^{\alpha}(t_{1},t_{2}) & =2D_{\alpha}\delta(t_{1}-t_{2})\nonumber \\
 & +\sum_{\beta}g_{\alpha\beta}^{2}\pint x\mu^{\beta}(x)\phi(x(t_{1}))\phi(x(t_{2})).\label{eq:C_eta_multipop}
\end{align}
Again, the rate function can be interpreted as a log-likelihood; its
derivative leads to (Appendix E1)
\begin{align}
\mathcal{S}_{\tau_{\alpha}\dot{x}+U_{\alpha}^{\prime}(x)}^{\alpha}(f) & =2D_{\alpha}+\sum_{\beta}g_{\alpha\beta}^{2}\mathcal{S}_{\phi(x)}^{\beta}(f),\label{eq:inference_condition_multipop}
\end{align}
which generalizes \prettyref{eq:inference_condition_stat} to multiple
populations.

Using \prettyref{eq:inference_condition_multipop}, the inferred connectivity
$g_{\alpha\beta}$ matches the ground truth well; accordingly, two
unconnected populations (\prettyref{fig:inference_multipop}\textbf{A},\textbf{B})
can be clearly distinguished from a more involved network where one
population ($\alpha=1$) is only active due to the recurrent input
from the other population ($\alpha=2$, \prettyref{fig:inference_multipop}\textbf{C},\textbf{D}).
The method can thus distinguish intrinsically generated activity from
a case where activity is driven from outside the network. However,
inference of a unique set of parameters is only possible if the output
spectra $\mathcal{S}_{\phi(x)}^{\alpha}(f)$ differ sufficiently across
$\alpha$. If the output spectra match closely, \prettyref{eq:inference_condition_multipop}
leads to a degenerate set of solutions that satisfy $\sum_{\beta}g_{\alpha\beta}^{2}=\mathrm{const.}$
and are all equally likely given the data (Appendix E2).

\paragraph{Discussion.--}

In this Letter, we found a tight link between the field theoretical
approach to neuronal networks and its counterpart based on large deviations
theory. We obtained the rate function of the empirical measure for
the widely used and analytically solvable model of a recurrent neuronal
network \citep{Sompolinsky88_259} by field-theoretical methods. This
rate function generalizes the seminal result by Ben Arous and Guionnet
\citep{Ben-Arous95_455,Guionnet97_183} to arbitrary potentials, transfer
functions, and multiple populations. Intriguingly, our derivation
elucidates that the rate function is identical to the effective action
and takes the form of a Kullback--Leibler divergence, akin to Sanov's
theorem for sums of i.i.d.~random variables \citep{Touchette09,Mezard_Montanari09}.
The rate function can thus be interpreted as a distance between an
empirical measure, for example given by data, and the activity statistics
of the network model. This result allows us to address the inverse
problem of inferring the parameters of the connectivity and external
input from a set of trajectories and to determine the potential and
the transfer function.

We here restricted the analysis to networks with
independently drawn random weights with zero mean. Since correlated
weights have a profound impact on the dynamics that can be captured
using both field theory \citep{Marti18_062314} and large deviations
theory \citep{Faugeras15_4701,Faugeras19_arxiv}, it is an interesting
challenge to extend the analysis in this direction. Likewise, synaptic
weights with non-vanishing mean, as they appear in sparsely-connected
networks, present an interesting extension, because they promote fluctuation-driven
states when feedback is sufficiently positive. Another important deviation
from independent weights in biological neural networks are motifs
\citep{Song05_0507}, which pose a significant challenge already for
the field-theoretical approach \citep{Dahmen20_365072v1}. Beyond
the weight statistics, we assumed that the dynamics are governed by
the first-order differential equation \eqref{eq:network_ode}. Indeed,
the field-theoretical approach can be generalized to a much broader
class of dynamics that do not necessarily possess an action \citep{Keup21_021064};
hence, it seems possible to also derive large deviations results for
more general dynamics. In this regard, the extension to spiking networks
is a particularly interesting but also challenging future direction.
Whether the model, \prettyref{eq:network_ode}, with its current limitations---the
independent weights and the first-order dynamics---allows accurate
inference of network parameters from cortical recordings is an intriguing
question for further research.

The unified description of random networks by statistical field theory
and large deviations theory opens the door to established techniques
from either domain to capture beyond mean-field behavior. Such corrections
are important for small or sparse networks with non--vanishing mean
connectivity, to explain correlated neuronal activity, and to study
information processing in finite-size networks with realistically
limited resources. We here make a first step by computing fluctuation
corrections from the rate function. The quantitative theory explains
near-critical fluctuations for $g\in[1,1+\delta(N)]$ and we discover
that expansive gain functions, as found in biology \citep{Roxin11_16217},
lead to qualitatively different collective behavior than the well-studied
contractive sigmoidal ones: The former feature meta--stable network
states with noise-induced first order transitions between them; the
latter allow for only a single solution and show second order phase
transitions.
\begin{acknowledgments}
We are grateful to Olivier Faugeras and Etienne Tanr\'e for helpful
discussions on LDT of neuronal networks, to Anno Kurth for pointing
us to the Fréchet derivative and to Alexandre Ren\'e, David Dahmen,
Kirsten Fischer, and Christian Keup for feedback on an earlier version
of the manuscript. This work was partly supported by the Helmholtz
young investigator's group VH-NG-1028, European Union Horizon 2020
grant 785907 (Human Brain Project SGA2) and the Human Frontier Science
Program RGP0057/2016 grant.
\end{acknowledgments}

\end{document}

\selectlanguage{english}%
\global\long\def\T{\!\mathsf{T}}%
\global\long\def\l{\langle}%
\global\long\def\r{\rangle}%
\global\long\def\ll{\langle\langle}%
\global\long\def\rr{\rangle\rangle}%
\global\long\def\tr{\mathrm{tr}}%
\global\long\def\lr#1{\left\langle #1\right\rangle }%
\global\long\def\vect#1{\bm{#1}}%
\global\long\def\matr#1{\bm{#1}}%
\global\long\def\uint#1{\int\mathrm{d}#1\,}%
\global\long\def\dint#1#2#3{\int_{#2}^{#3}\mathrm{d}#1\,}%
\global\long\def\pint#1{\int\mathcal{D}#1\,}%
\global\long\def\C{C}%
\global\long\def\Ct{\tilde{C}}%
\global\long\def\k{\ell}%
\global\long\def\evalat#1{\left.#1\right|}%
\global\long\def\xt{\tilde{x}}%
\global\long\def\erf{\mathrm{erf}}%
\global\long\def\clip{\mathrm{clip}}%

\title{Large Deviations Approach to Random Recurrent Neuronal Networks:\\
Parameter Inference and Fluctuation--Induced Transitions\\
(Supplemental Material)}
\author{Alexander van Meegen}
\affiliation{Institute of Neuroscience and Medicine (INM-6) and Institute for Advanced
Simulation (IAS-6) and JARA-Institute Brain Structure-Function Relationships
(INM-10), Jülich Research Centre, Jülich, Germany}
\affiliation{Institute of Zoology, University of Cologne, 50674 Cologne, Germany}
\author{Tobias Kühn}
\affiliation{Institute of Neuroscience and Medicine (INM-6) and Institute for Advanced
Simulation (IAS-6) and JARA-Institute Brain Structure-Function Relationships
(INM-10), Jülich Research Centre, Jülich, Germany}
\affiliation{Department of Physics, Faculty 1, RWTH Aachen University, Aachen,
Germany}
\affiliation{Laboratoire de Physique de l'ENS, Laboratoire MSC de l'Université
de Paris, CNRS, Paris, France}
\author{Moritz Helias}
\affiliation{Institute of Neuroscience and Medicine (INM-6) and Institute for Advanced
Simulation (IAS-6) and JARA-Institute Brain Structure-Function Relationships
(INM-10), Jülich Research Centre, Jülich, Germany}
\affiliation{Department of Physics, Faculty 1, RWTH Aachen University, Aachen,
Germany}
\date{\today}

\maketitle
\selectlanguage{american}%
\tableofcontents{}

\subsection{Rate Function (Single Population)}

\subsubsection{Scaled Cumulant Generating Functional}

Here, we derive the scaled cumulant generating functional and the
saddle-point equations. The first steps of the derivations are akin
to the manipulations presented in \citep{Schuecker16b_arxiv,Schuecker18_041029},
thus we keep the presentation concise. We interpret the stochastic
differential equations governing the network dynamics in the Itô convention.
Using the Martin--Siggia--Rose--de Dominicis--Janssen path integral
formalism, the expectation $\lr{\cdot}_{\vect x|\matr J}$ of some
arbitrary functional $G(\vect x)$ can be written as
\begin{align*}
\l\lr{G(\vect x)}_{\vect x|\matr J,\vect{\xi}}\r_{\vect{\xi}} & =\pint{\vect x}\lr{\delta(\dot{\vect x}+U^{\prime}(\vect x)+\matr J\phi(\vect x)+\vect{\xi})}_{\vect{\xi}}G(\vect x)\\
 & =\pint{\vect x}\pint{\vect{\tilde{x}}}\,e^{S_{0}(\vect x,\vect{\tilde{x}})-\vect{\tilde{x}}^{\T}\matr J\phi(\vect x)}G(\vect x),
\end{align*}
where we used the Fourier representation $\delta(x)=\frac{1}{2\pi i}\int_{-i\infty}^{i\infty}e^{\tilde{x}x}d\tilde{x}$
in every timestep in the second step and defined the action
\begin{align*}
S_{0}(\vect x,\vect{\tilde{x}}) & =\vect{\tilde{x}}^{\T}(\dot{\vect x}+U^{\prime}(\vect x))+D\vect{\tilde{x}}^{\T}\vect{\tilde{x}}.
\end{align*}
An additional average over realizations of the connectivity $\matr J\stackrel{\text{i.i.d.}}{\sim}\mathcal{N}(0,N^{-1}g^{2})$
only affects the term $-\vect{\tilde{x}}^{\T}\matr J\phi(\vect x)$
in the action and results in 
\begin{align*}
\l e^{-\vect{\tilde{x}}^{\T}\matr J\phi(\vect x)}\r_{\matr J} & =\pint C\pint{\tilde{C}}e^{-N\,C^{\T}\tilde{C}+\frac{g^{2}}{2}\vect{\tilde{x}}^{\T}C\vect{\tilde{x}}+\phi(\vect x)^{\T}\tilde{C}\phi(\vect x)},
\end{align*}
where we introduced the network--averaged auxiliary field
\[
C(u,v)=\frac{1}{N}\sum_{i=1}^{N}\phi(x_{i}(u))\phi(x_{i}(v))
\]
via a Hubbard--Stratonovich transformation. The average over the
connectivity and the subsequent Hubbard--Stratonovich transformation
decouple the dynamics across units; afterwards the units are only
coupled through the global fields $C$ and $\tilde{C}$.

Now, we consider the scaled cumulant generating functional of the
empirical density
\begin{align*}
W_{N}(\ell) & =\frac{1}{N}\ln\lr{\lr{e^{\sum_{i=1}^{N}\k(x_{i})}}_{\vect x|\matr J}}_{\matr J}.
\end{align*}
Using the above results and the abbreviation $\phi(x)\equiv\phi$,
it can be written as
\begin{align*}
W_{N}(\ell) & =\frac{1}{N}\ln\,\pint C\pint{\tilde{C}}e^{-N\,C^{\T}\tilde{C}+N\,\Omega_{\ell}(C,\tilde{C})},\\
\Omega_{\ell}(C,\tilde{C}) & =\ln\,\pint x\pint{\xt}e^{S_{0}(x,\xt)+\frac{g^{2}}{2}\xt^{\T}C\xt+\phi^{\T}\tilde{C}\phi+\k(x)},
\end{align*}
where the $N$ in front of the single--particle cumulant generating
functional $\Omega$ results from the factorization of the $N$ integrals
over $x_{i}$ and $\tilde{x}_{i}$ each; thus it is a hallmark of
the decoupled dynamics. Next, we approximate the $C$ and $\tilde{C}$
integrals in a saddle--point approximation which yields
\begin{align*}
W_{N}(\ell) & =-C_{\ell}^{\T}\tilde{C}_{\ell}+\Omega_{\ell}(C_{\ell},\tilde{C}_{\ell})+O(\ln(N)/N),
\end{align*}
where $C_{\ell}$ and $\tilde{C}_{\ell}$ are determined by the saddle--point
equations
\begin{align*}
\C_{\k} & =\evalat{\partial_{\Ct}\Omega_{\k}(\C,\Ct)}_{\C_{\k},\Ct_{\k}},\\
\Ct_{\k} & =\evalat{\partial_{\C}\Omega_{\k}(\C,\Ct)}_{\C_{\k},\Ct_{\k}}.
\end{align*}
Here, $\partial_{C}$ denotes a partial functional derivative. In
the limit $N\to\infty$, the remainder $O(\ln(N)/N)$ vanishes and
the saddle--point approximation becomes exact.

\subsubsection{Rate Function}

Here, we derive the rate function from the scaled cumulant generating
functional. According to the Gärtner-Ellis theorem \citep{Touchette09},
we obtain the rate function via the Legendre transformation
\begin{align}
H(\mu) & =\pint x\mu(x)\k_{\mu}(x)-W_{\infty}(\k_{\mu})\label{eq:def_H_Legendre}
\end{align}
with $\k_{\mu}$ implicitly defined by 
\begin{align}
\mu & =W_{\infty}^{\prime}(\ell_{\mu}).\label{eq:implicit_eq_ell_mu}
\end{align}
Using the Gärtner-Ellis theorem, we implicitly assume that $H(\mu)$
is convex \citep{Touchette09}. This is, however, not the same as
assuming that $\mu$, or the most likely empirical measure $\bar{\mu}$,
is concave. The latter would be a serious restriction as it would
prohibit for example treating the bistable case we investigate in
the manuscript. A concave $P(\mu)$, and hence a convex $H(\mu)$,
simply corresponds to the situation with a single most likely measure
$\bar{\mu}$ but it does not put restrictions on $\bar{\mu}$ itself.
In particular, $\bar{\mu}$ may still be bimodal.

Due to the saddle--point equations, the derivative of the cumulant
generating functional in \prettyref{eq:implicit_eq_ell_mu} simplifies
to $W_{\infty}^{\prime}(\ell_{\mu})=\evalat{(\partial_{\ell}\Omega_{\k})(\C_{\ell},\Ct_{\ell})}_{\ell_{\mu}}$
where the derivative only acts on the $\ell$ that is explicit in
$\Omega_{\ell}(C_{\ell},\tilde{C_{\ell}})$ and not on the implicit
dependencies through $C_{\ell}$, $\tilde{C_{\ell}}$. Thus, \prettyref{eq:implicit_eq_ell_mu}
yields
\begin{align*}
\mu(x) & =\frac{\pint{\xt}e^{S_{0}(x,\xt)+\frac{g^{2}}{2}\xt^{\T}C_{\ell_{\mu}}\xt+\phi^{\T}\tilde{C}_{\ell_{\mu}}\phi+\k_{\mu}(x)}}{\pint x\pint{\xt}e^{S_{0}(x,\xt)+\frac{g^{2}}{2}\xt^{\T}C_{\ell_{\mu}}\xt+\phi^{\T}\tilde{C}_{\ell_{\mu}}\phi+\k_{\mu}(x)}}.
\end{align*}
Taking the logarithm and using $W_{\infty}(\ell_{\mu})+C_{\ell_{\mu}}^{\T}\tilde{C}_{\ell_{\mu}}=\Omega_{\ell_{\mu}}(C_{\ell_{\mu}},\tilde{C}_{\ell_{\mu}})$
leads to
\begin{align*}
\k_{\mu}(x)= & \ln\frac{\mu(x)}{\pint{\xt}e^{S_{0}(x,\xt)+\frac{g^{2}}{2}\xt^{\T}C_{\ell_{\mu}}\xt}}+W_{\infty}(\ell_{\mu})+C_{\ell_{\mu}}^{\T}\tilde{C}_{\ell_{\mu}}-\phi^{\T}\tilde{C}_{\ell_{\mu}}\phi.
\end{align*}
Inserting $\k_{\mu}(x)$ into the Legendre transformation \eqref{eq:def_H_Legendre}
yields
\begin{align*}
H(\mu)= & \pint x\mu(x)\ln\frac{\mu(x)}{\pint{\xt}e^{S_{0}(x,\xt)+\frac{g^{2}}{2}\xt^{\T}C_{\ell_{\mu}}\xt}}+C_{\ell_{\mu}}^{\T}\tilde{C}_{\ell_{\mu}}-C_{\mu}^{\T}\tilde{C}_{\ell_{\mu}}
\end{align*}
with
\begin{align*}
C_{\mu}(u,v) & =\pint x\mu(x)\phi(x(u))\phi(x(v)).
\end{align*}
Identifying $\mu(x)$ in the saddle--point equation
\begin{align*}
\C_{\k_{\mu}} & =\evalat{\partial_{\Ct}\Omega_{\k}(\C,\Ct)}_{\C_{\k_{\mu}},\Ct_{\k_{\mu}}}=\frac{\pint x\pint{\xt}\phi\phi e^{S_{0}(x,\xt)+\frac{g^{2}}{2}\xt^{\T}C_{\ell_{\mu}}\xt+\phi^{\T}\tilde{C}_{\ell_{\mu}}\phi+\k_{\mu}(x)}}{\pint x\pint{\xt}e^{S_{0}(x,\xt)+\frac{g^{2}}{2}\xt^{\T}C_{\ell_{\mu}}\xt+\phi^{\T}\tilde{C}_{\ell_{\mu}}\phi+\k_{\mu}(x)}}
\end{align*}
yields
\begin{align*}
\C_{\k_{\mu}}(u,v) & =\pint x\mu(x)\phi(x(u))\phi(x(v))
\end{align*}
and thus $\C_{\k_{\mu}}=C_{\mu}$. Accordingly, the last two terms
in the Legendre transformation cancel and we arrive at
\begin{equation}
H(\mu)=\pint x\mu(x)\ln\frac{\mu(x)}{\pint{\xt}e^{S_{0}(x,\xt)+\frac{g^{2}}{2}\xt^{\T}C_{\mu}\xt}}\label{eq:rate_fct_appendix}
\end{equation}
where still $C_{\mu}(u,v)=\pint x\mu(x)\phi(x(u))\phi(x(v))$.

In the main text, we use the notation
\begin{align*}
\pint{\xt}e^{S_{0}(x,\xt)+\frac{g^{2}}{2}\xt^{\T}C_{\mu}\xt} & =\lr{\delta(\dot{x}+U^{\prime}(x)-\eta)}_{\eta}
\end{align*}
with $C_{\eta}=2D\delta+g^{2}C_{\mu}$ appearing in the rate function.
Indeed, using the Martin--Siggia--Rose--de Dominicis--Janssen
formalism, we have
\begin{align*}
\lr{\delta(\dot{x}+U^{\prime}(x)-\eta)}_{\eta} & =\pint{\xt}e^{\tilde{x}^{\T}(\dot{x}+U^{\prime}(x))}\langle e^{\tilde{x}^{\T}\eta}\rangle_{\eta}\\
 & =\pint{\xt}e^{\tilde{x}^{\T}(\dot{x}+U^{\prime}(x))+\frac{1}{2}\xt^{\T}C_{\eta}\xt},
\end{align*}
which shows that the two notations are equivalent since $\tilde{x}^{\T}(\dot{x}+U^{\prime}(x))+\frac{1}{2}\xt^{\T}C_{\eta}\xt=S_{0}(x,\xt)+\frac{g^{2}}{2}\xt^{\T}C_{\mu}\xt$
for $C_{\eta}=2D\delta+g^{2}C_{\mu}$.

\subsubsection{Equivalence to Ben Arous and Guionnet (1995)}

Here, we show explicitly that the rate function we obtained generalizes
the rate function obtained by Ben Arous and Guionnet \citep{Ben-Arous95_455},
whose limitation to finite temperature and time was lifted later \citep{Guionnet97_183}.
We start with Theorem 4.1 in \citep{Ben-Arous95_455} adapted to our
notation: Define
\begin{align*}
Q(x) & :=\pint{\xt}e^{\tilde{x}^{\T}(\dot{x}+U^{\prime}(x))+\frac{1}{2}\xt^{\T}\xt}
\end{align*}
and
\begin{align*}
G(\mu) & :=\pint x\mu(x)\,\ln\left(\langle e^{gy^{\T}(\dot{x}+U^{\prime}(x))-\frac{g^{2}}{2}y^{\T}y}\rangle_{y}\right),
\end{align*}
where $\langle\cdot\rangle_{y}$ is the expectation value over a zero--mean
Gaussian process $y$ with $C_{\mu}(u,v)=\pint x\mu(x)x(u)x(v)$,
written as $\langle\cdot\rangle_{y}=\pint y\pint{\tilde{y}}\left(\cdot\right)\,e^{\tilde{y}^{\T}y+\frac{1}{2}\tilde{y}^{\T}C_{\mu}\tilde{y}}$.
With the Kullback--Leibler divergence $D_{\text{KL}}(\mu\,|\,Q)$,
Theorem 4.1 states that the function
\begin{align*}
\tilde{H}(\mu) & =\begin{cases}
D_{\text{KL}}(\mu\,|\,Q)-G(\mu) & \text{if }D_{\text{KL}}(\mu\,|\,Q)<\infty\\
+\infty & \text{otherwise}
\end{cases}
\end{align*}
is a good rate function.

Now we relate $\tilde{H}$ to the rate function that is derived above,
\prettyref{eq:rate_fct_appendix}. Using the Onsager--Machlup action,
we can write
\begin{align*}
D_{\text{KL}}(\mu\,|\,Q) & =\pint x\mu(x)\ln\frac{\mu(x)}{e^{-S_{\mathrm{OM}}(x)}}+\mathcal{C}
\end{align*}
with $S_{\mathrm{OM}}(x)=\frac{1}{2}(\dot{x}+U^{\prime}(x))^{\T}(\dot{x}+U^{\prime}(x))$.
Next, we transform $gy\to y$, $\tilde{y}/g\to\tilde{y}$ and solve
the integral over $y$ in $G(\mu)$:
\begin{align*}
\pint ye^{-\frac{1}{2}y^{\T}y+y^{\T}(\dot{x}+U^{\prime}(x)+\tilde{y})} & \propto e^{S_{\mathrm{OM}}[x]+\tilde{y}^{\T}(\dot{x}+U^{\prime}(x))+\frac{1}{2}\tilde{y}^{\T}\tilde{y}}.
\end{align*}
The Onsager--Machlup action in the logarithm in $D_{\text{KL}}(\mu\,|\,Q)$
and $G(\mu)$ cancel and we arrive at
\begin{align*}
\tilde{H}(\mu) & =\pint x\mu(x)\ln\frac{\mu(x)}{\pint{\tilde{y}}e^{\tilde{y}^{\T}(\dot{x}+U^{\prime}(x))+\frac{1}{2}\tilde{y}^{\T}(g^{2}C_{\mu}+\delta)\tilde{y}}}
\end{align*}
up to an additive constant that we set to zero. Since $C_{\mu}(u,v)=\pint x\mu(x)x(u)x(v)$,
the rate function by Ben Arous and Guionnet is thus equivalent to
\prettyref{eq:rate_fct_appendix} with $\phi(x)=x$ and $D=\frac{1}{2}$.

\subsubsection{Background on Rate Function}

\paragraph{Relation to Sompolinsky, Crisanti, Sommers (1988)}

Here, we relate the approach that we laid out in the main text to
the approach pioneered by Sompolinsky, Crisanti, and Sommers \citep{Sompolinsky88_259}
(reviewed in \citep{Schuecker18_041029,Crisanti18_062120}) using
our notation for consistency. Therein, the starting point is the scaled
cumulant--generating functional

\begin{align*}
\hat{W}_{N}(j) & =\frac{1}{N}\ln\lr{\lr{e^{j^{\T}\vect x}}_{\vect x|\matr J}}_{\matr J},
\end{align*}
which gives rise to the cumulants of the trajectories. For the linear
functional
\begin{align*}
\ell(x) & =j^{\T}x,
\end{align*}
we have $\sum_{i=1}^{N}\k(x_{i})=j^{\T}\vect x$ and thus $W_{N}(j^{\T}x)=\hat{W}_{N}(j)$.
Put differently, the scaled cumulant--generating functional of the
trajectories $\hat{W}_{N}(j)$ is a special case of the more general
scaled cumulant--generating functional $W_{N}(\ell)$ we consider
in this manuscript. Of course one can start from the scaled cumulant--generating
functional of the observable of interest and derive the corresponding
rate function. Conversely, we show below how to obtain the rate function
of a specific observable from the rate function of the empirical measure.

\paragraph{Contraction Principle}

Here, we relate the rather general rate function of the empirical
measure $H(\mu)$ to the rate function of a particular observable
$I(C)$. As an example, we choose the correlation function
\begin{align*}
C(u,v) & =\frac{1}{N}\sum_{i=1}^{N}\phi(x_{i}(u))\phi(x_{i}(v))
\end{align*}
because it is a quantity that arises naturally during the Hubbard--Stratonovich
transformation. The generic approach to this problem is given by the
contraction principle \citep{Touchette09}:
\begin{align*}
I(C) & =\inf_{\mu\:\text{s.t.}\:C=\pint x\mu(x)\phi\phi}H(\mu).
\end{align*}
Here, the infimum is constrained to the empirical measures that give
rise to the correlation function $C$, i.e.~those that fulfill $C(u,v)=\pint x\mu(x)\phi(x(u))\phi(x(v))$.
Writing $H(\mu)$ as the Legendre transform of the scaled cumulant--generating
functional, $H(\mu)=\inf_{\ell}[\pint x\mu(x)\ell(x)-W_{\infty}(\k)${]},
the empirical measure only appears linearly. Using a Lagrange multiplier
$k(u,v)$, the infimum over $\mu$ leads to the constraint $\ell(x)=\phi^{\T}k\phi$
and we arrive at
\begin{align*}
I(C) & =\inf_{k}[k^{\T}C-W_{\infty}(\phi^{\T}k\phi)].
\end{align*}
Once again, we see how to relate $W_{N}(\ell)$ to a specific observable---this
time for the choice $\ell(x)=\phi^{\T}k\phi$.

Up to this point, the discussion applies to any observable. For the
current example, we can proceed a bit further. With the redefinition
$\tilde{C}+k\to\tilde{C}$, we get
\begin{align*}
W_{\infty}(\phi^{\T}k\phi) & =\mathrm{extr}_{C,\tilde{C}}\left[-\C^{\T}\Ct+\C^{\T}k+\Omega_{0}(\C,\Ct)\right],\\
\Omega_{0}(C,\tilde{C}) & =\ln\,\pint x\pint{\xt}e^{S_{0}(x,\xt)+\frac{g^{2}}{2}\xt^{\T}C\xt+\phi^{\T}\tilde{C}\phi},
\end{align*}
which made $\Omega_{0}$ independent of $k$. Now we can take the
infimum over $k$, leading to
\begin{align}
I(C) & =\mathrm{extr}_{\tilde{C}}\left[C^{\T}\Ct-\Omega_{0}(C,\Ct)\right].\label{eq:Rate_fct_C_contraction_principle}
\end{align}
The remaining extremum gives rise to the condition
\begin{align*}
C & =\frac{\pint x\pint{\xt}\phi\phi e^{S_{0}(x,\xt)+\frac{g^{2}}{2}\xt^{\T}C_{\phi}\xt+\phi^{\T}\tilde{C}\phi}}{\pint x\pint{\xt}e^{S_{0}(x,\xt)+\frac{g^{2}}{2}\xt^{\T}C_{\phi}\xt+\phi^{\T}\tilde{C}\phi}},
\end{align*}
i.e.~a self--consistency condition for the correlation function.

As a side remark, we mention that the expression in the brackets of
\prettyref{eq:Rate_fct_C_contraction_principle} is the joint effective
action for $C$ and $\tilde{C}$, because for $N\rightarrow\infty$,
the action equals the effective action. This result is therefore analogous
to the finding that the effective action in the Onsager--Machlup
formalism is given as the extremum of its counterpart in the Martin--Siggia--Rose--de
Dominicis--Janssen formalism \citep[Eq.(24)]{Stapmanns20_042124}.
The only difference is that here, we are dealing with second order
statistics and not just mean values. The origin of this finding is
the same in both cases: we are only interested in the statistics of
the physical quantity (the one without tilde, $x$ or $C$, respectively).
Therefore we only introduce a source field ($k$ in the present case)
for this one, but not for the auxiliary field, which amounts to setting
the source field of the latter to zero. This is translated into the
extremum in \prettyref{eq:Rate_fct_C_contraction_principle} over
the auxiliary variable \citep[Appendix 5]{Stapmanns20_042124}.

\paragraph{Tail Probability}

Large deviations results are often stated for the tail probability
$\mathbb{P}(x>\theta)$ where $\theta$ is in the tail. Since the
notion of a tail cannot be unambiguously defined for quantities like
the empirical measure or correlation functions, at least not in an
obvious way, we here give an example how to relate the rate function
of the empirical measure to a tail probability.

First, we use the contraction principle to get a rate function for
a scalar quantity, e.g. the order parameter $q=\pint x\mu(x)\phi(x(t))\phi(x(t))$
where $t$ is large but fixed such that the measure becomes stationary:
\begin{align*}
I(q) & =\inf_{\mu\:\text{s.t.}\:q=\pint x\mu(x)\phi\phi}H(\mu).
\end{align*}
Since $q$ is a scalar quantity, one obtains the tail probability
as $\ln\mathbb{P}(q>\theta)\simeq-NI(q=\theta)$.

Below, we calculate both the mean and the variance of $q$. In general,
this would not be sufficient to obtain a tail estimate. However, the
numerics indicate that the tail is indeed Gaussian (\prettyref{fig:order_param_fluct}\textbf{D})
such that the first two cumulants are indeed sufficient.

\subsection{Inference \& Prediction (Single Population)}

\subsubsection{Log--Likelihood Derivative}

Here, we calculate the derivatives of the log--likelihood with respect
to the parameters $g$ and $D$. In terms of the rate function, we
have
\begin{align*}
\partial_{a}\ln P(\mu\,|\,g,D) & \simeq-N\partial_{a}H(\mu\,|\,g,D)
\end{align*}
where $a$ denotes either $g$ or $D$. The parameters appear only
in the cross entropy
\begin{align*}
\partial_{a}H(\mu) & =-\pint x\mu(x)\partial_{a}\ln\lr{\delta(\dot{x}+U^{\prime}(x)-\eta)}_{\eta}
\end{align*}
through the correlation function $C_{\eta}(u,v)=2D\delta(u-v)+g^{2}\negthinspace\pint x\mu(x)\phi(x(u))\phi(x(v))$.
Above, we showed that
\begin{align*}
\lr{\delta(\dot{x}+U^{\prime}(x)-\eta)}_{\eta} & =\pint{\xt}e^{\tilde{x}^{\T}(\dot{x}+U^{\prime}(x))+\frac{1}{2}\xt^{\T}C_{\eta}\xt}.
\end{align*}
Because $\tilde{x}$ is at most quadratic in the exponent, the integral
is solvable and we get
\begin{align*}
\lr{\delta(\dot{x}+U^{\prime}(x)-\eta)}_{\eta} & =\frac{e^{-\frac{1}{2}(\dot{x}+U^{\prime}(x))^{\T}C_{\eta}^{-1}(\dot{x}+U^{\prime}(x))}}{\sqrt{\det(2\pi C_{\eta})}}.
\end{align*}
Note that the normalization $1/\sqrt{\det(2\pi C_{\eta})}$ does not
depend on the potential $U$. Now we can take the derivatives of $\ln\lr{\delta(\dot{x}+U^{\prime}(x)-\eta)}_{\eta}$
and get
\begin{align*}
\partial_{a}\ln\lr{\delta(\dot{x}+U^{\prime}(x)-\eta)}_{\eta} & =-\frac{1}{2}(\dot{x}+U^{\prime}(x))^{\T}\frac{\partial C_{\eta}^{-1}}{\partial a}(\dot{x}+U^{\prime}(x))-\frac{1}{2}\partial_{a}\tr\ln C_{\eta}
\end{align*}
where we used $\ln\det C=\tr\ln C$. With this, we arrive at
\begin{align*}
\partial_{a}H(\mu) & =\frac{1}{2}\tr\left(C_{0}\frac{\partial C_{\eta}^{-1}}{\partial a}\right)+\frac{1}{2}\tr\left(\frac{\partial C_{\eta}}{\partial a}C_{\eta}^{-1}\right)
\end{align*}
where the integral over the empirical measure gave rise to $C_{0}=\pint x\mu(x)(\dot{x}+U^{\prime}(x))(\dot{x}+U^{\prime}(x))$
and we used $\partial_{a}\ln C=\frac{\partial C}{\partial a}C^{-1}$.
Finally, using $\frac{\partial C}{\partial a}C^{-1}=CC^{-1}\frac{\partial C}{\partial a}C^{-1}=-C\frac{\partial C^{-1}}{\partial a}$,
we get
\begin{align*}
\partial_{a}\ln P(\mu\,|\,g,D) & \simeq-\frac{N}{2}\tr\left((C_{0}-C_{\eta})\frac{\partial C_{\eta}^{-1}}{\partial a}\right)
\end{align*}
as stated in the main text.

The derivative vanishes for $C_{0}=C_{\eta}$. Assuming stationarity,
in Fourier domain this condition reads
\begin{align}
\mathcal{S}_{\dot{x}+U^{\prime}(x)}(f) & =2D+g^{2}\mathcal{S}_{\phi(x)}(f),\label{eq:inference_condition_stat}
\end{align}
where $\mathcal{S}_{X}(f)$ denotes the network--averaged power spectrum
of the observable $X$.

\subsubsection{Model Comparison}

\begin{figure}
\includegraphics{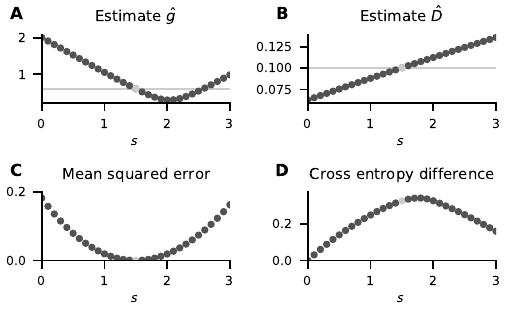}\caption{Model comparison for $\phi(x)=\protect\erf(\sqrt{\pi}x/2)$ and $U(x)=\frac{1}{2}x^{2}-s\ln\cosh x$.
\textbf{A},\textbf{B} Maximum likelihood estimates of $\hat{g}$ and
$\hat{D}$ for given choices of $s$. True values of $g$ and $D$
indicated as gray lines; estimates at the true value $s=1.5$ indicated
as gray symbols. \textbf{C} Mean squared error between left-- and
right--hand--side of \prettyref{eq:inference_condition_stat} for
given $s$. \textbf{D} Cross entropy difference between model with
$s=0$ and with given $s$. Further parameters as in Fig.~1 in the
main text.\label{fig:model_comparison}}
\end{figure}

Parameter estimation allows us to determine the statistical properties
of the recurrent connectivity $g$ and the external input $D$. However,
this leaves the potential $U$ and the transfer function $\phi$ unspecified.
Here we determine $U$ and $\phi$ using model comparison techniques
\citep{Mackay2003}.

We consider two options to obtain $U$ and $\phi$: comparing the
mean squared error in \prettyref{eq:inference_condition_stat} for
the inferred parameters and comparing the likelihood of the inferred
parameters. For the latter option, we can use the rate function from
\prettyref{eq:rate_fct_appendix}. Given two choices $U_{i}$, $\phi_{i}$,
$i\in\{1,2\}$, with corresponding inferred parameters $\hat{g}_{i}$,
$\hat{D}_{i}$, we have
\begin{align}
\ln\frac{P(\mu\,|\,U_{1},\phi_{1},\hat{g}_{1},\hat{D}_{1})}{P(\mu\,|\,U_{2},\phi_{2},\hat{g}_{2},\hat{D}_{2})} & \simeq-N(H_{1}-H_{2})\label{eq:likelihood_ratio}
\end{align}
with $H_{i}\equiv H(\mu\,|\,U_{i},\phi_{i},\hat{g}_{i},\hat{D}_{i})$.
The difference $H_{1}-H_{2}$ equals the difference of the minimal
cross entropies for the respective choices $U_{i}$, $\phi_{i}$.
Assuming an infinite observation time, this difference can be expressed
as an integral that is straightforward to evaluate numerically (see
below).

To illustrate the procedure, we consider the potential
\begin{align*}
U(x) & =\frac{1}{2}x^{2}-s\ln\cosh x,
\end{align*}
which is bistable for $s>1$ \citep{Stern14_062710} and determine
$s$ using the mean squared error and the cross entropy difference
(see \prettyref{fig:model_comparison}). Parameter estimation yields
estimates $\hat{g}$ and $\hat{D}$ that depend on $s$ (\prettyref{fig:model_comparison}\textbf{A},\textbf{B}).
The mean squared error displays a clear minimum at the true value
$s=1.5$ (\prettyref{fig:model_comparison}\textbf{C}) whereas the
maximal cross entropy occurs at a value larger than $s=1.5$ (\prettyref{fig:model_comparison}\textbf{D}).
The latter effect arises because the cross entropy is dominated by
the parameter estimates, thus the mean squared error provides a more
reliable criterion in this case.

\paragraph{Cross Entropy Difference}

Here, we express the cross entropy difference
\begin{align*}
H_{1}-H_{2} & :=H(\mu\,|\,U_{1},\phi_{1},\hat{g}_{1},\hat{D}_{1})-H(\mu\,|\,U_{2},\phi_{2},\hat{g}_{2},\hat{D}_{2})
\end{align*}
in a form that can be evaluated numerically. Using the rate function,
we get
\[
H_{1}-H_{2}=\pint x\mu(x)\ln\frac{\lr{\delta(\dot{x}+U_{2}^{\prime}(x)-\eta_{2})}_{\eta_{2}}}{\lr{\delta(\dot{x}+U_{1}^{\prime}(x)-\eta_{1})}_{\eta_{1}}}
\]
with $C_{\eta_{i}}=2\hat{D}_{i}\delta+\hat{g}_{i}^{2}\pint x\mu(x)\phi_{i}\phi_{i}$.
Again, we use
\begin{align*}
\lr{\delta(\dot{x}+U^{\prime}(x)-\eta)}_{\eta} & =\frac{e^{-\frac{1}{2}(\dot{x}+U^{\prime}(x))^{\T}C_{\eta}^{-1}(\dot{x}+U^{\prime}(x))}}{\sqrt{\det(2\pi C_{\eta})}}
\end{align*}
to arrive at
\begin{align*}
H_{1}-H_{2}= & -\frac{1}{2}\tr\left(C_{1}C_{\eta_{1}}^{-1}\right)-\frac{1}{2}\tr\ln C_{\eta_{1}}+\frac{1}{2}\tr\left(C_{2}C_{\eta_{2}}^{-1}\right)+\frac{1}{2}\tr\ln C_{\eta_{2}}
\end{align*}
with $C_{i}=\pint x\mu(x)(\dot{x}+U_{i}^{\prime}(x))(\dot{x}+U_{i}^{\prime}(x))$.
For stationary correlation functions over infinite time intervals,
we can evaluate the traces as integrals over the power spectra:
\begin{align*}
\tr(AB^{-1}) & \propto\int_{-\infty}^{\infty}\frac{\tilde{A}(f)}{\tilde{B}(f)}df,\\
\tr\ln A & \propto\int_{-\infty}^{\infty}\ln(\tilde{A}(f))df.
\end{align*}
With this, we get
\begin{align*}
H_{1}-H_{2}\propto & -\frac{1}{2}\int_{-\infty}^{\infty}\frac{\mathcal{S}_{\dot{x}+U_{1}^{\prime}(x)}(f)}{2\hat{D}_{1}+\hat{g}_{1}^{2}\mathcal{S}_{\phi_{1}(x)}(f)}df-\frac{1}{2}\int_{-\infty}^{\infty}\ln(2\hat{D}_{1}+\hat{g}_{1}^{2}\mathcal{S}_{\phi_{1}(x)}(f))df\\
 & +\frac{1}{2}\int_{-\infty}^{\infty}\frac{\mathcal{S}_{\dot{x}+U_{2}^{\prime}(x)}(f)}{2\hat{D}_{2}+\hat{g}_{2}^{2}\mathcal{S}_{\phi_{2}(x)}(f)}df+\frac{1}{2}\int_{-\infty}^{\infty}\ln(2\hat{D}_{2}+\hat{g}_{2}^{2}\mathcal{S}_{\phi_{2}(x)}(f))df.
\end{align*}
Accordingly, the cross entropy difference can be evaluated with integrals
over the respective power spectra that can be obtained using Fast
Fourier Transformation.

\subsubsection{Activity Prediction}

\begin{figure}
\includegraphics{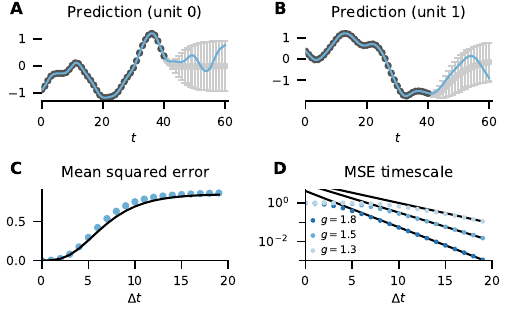}

\caption{Prediction of the future single--unit activity. \textbf{A},\textbf{B}
Prediction $\hat{x}$ with uncertainty $\sigma_{\hat{x}}$ (light
symbols) for two arbitrary units. Training data (dark symbols) determined
by the true trajectory (solid curve). \textbf{C} Network--averaged
mean squared error $\epsilon$ (symbols) and predicted uncertainty
$\sigma_{\hat{x}}^{2}$ (solid curve). \textbf{D} The error increases
on half of the timescale of the autocorrelation function: $(C_{x}(0)-\sigma_{\hat{x}}^{2})/C_{x}(0)$
(symbols) decreases asymptotically as $\mathcal{C}\exp(-2\Delta t/\tau_{c})$
(lines). Network parameters $\phi(x)=\protect\erf(\sqrt{\pi}x/2)$,
$U(x)=\frac{1}{2}x^{2}$, and $D=0$; further parameters as in Fig.~1
in the main text. \label{fig:prediction}}
\end{figure}

If the potential of the model is quadratic, $U(x)\propto\frac{1}{2}x^{2}$,
the measure $\bar{\mu}$ that minimizes the rate function corresponds
to a Gaussian process. For Gaussian processes, it is possible to perform
Bayes--optimal prediction only based on its correlation function
\citep{Matheron63_1246,Mackay2003}. Denoting the correlation function
of the process as $C_{x}$ (Appendix B4), the prediction is given
by
\begin{align}
\hat{x} & =\vect k^{\T}\vect K^{-1}\vect x\label{eq:pred_mean}
\end{align}
with $K_{ij}=C_{x}(t_{i},t_{j})$, $k_{i}=C_{x}(t_{i},\hat{t})$,
and $x_{i}=x(t_{i})$. Here $\hat{t}$ denotes the time point of the
prediction and $\{t_{i}\}$ a set of time points where the activity
is known. The predicted value $\hat{x}$ itself is Gaussian distributed
with variance
\begin{align}
\sigma_{\hat{x}}^{2} & =\kappa-\vect k^{\T}\vect K^{-1}\vect k\label{eq:pred_var}
\end{align}
where $\kappa=C_{x}(\hat{t},\hat{t})$. The variance $\sigma_{\hat{x}}^{2}$
quantifies the uncertainty associated with the prediction $\hat{x}$.

We use the self-consistent autocorrelation function from \prettyref{eq:rate_fct_appendix}
to predict the future activity of two arbitrary units using \prettyref{eq:pred_mean}
and \prettyref{eq:pred_var} (\prettyref{fig:prediction}\textbf{A},\textbf{B}).
The network--averaged mean squared error $\epsilon=\frac{1}{N}\sum_{i=1}^{N}(\hat{x}_{i}-x_{i})^{2}$
is well predicted by \prettyref{eq:pred_var} as shown in \prettyref{fig:prediction}\textbf{C}.
The timescale of the error is half of the timescale of the autocorrelation
function (Appendix B5). We plot $(C_{x}(0)-\sigma_{\hat{x}}^{2})/C_{x}(0)$
against an exponential decay $\mathcal{C}\exp(-2\tau/\tau_{c})$,
where $C_{x}(\tau)/C_{x}(0)\sim\exp(-\tau/\tau_{c})$, and find a
very good agreement (\prettyref{fig:prediction}\textbf{D}). Since
$\tau_{c}$ diverges for $g\searrow1$ (cf.~\citep{Sompolinsky88_259}),
the timescale of the error diverges as well.

\subsubsection{Self--Consistent Correlation Function}

Here, we describe how the self--consistent correlation function can
be obtained efficiently for quadratic single--unit potentials $U(x)=\frac{1}{2}x^{2}$.
The first part is a brief recapitulation of the approach in \citep{Sompolinsky88_259,Schuecker18_041029},
the second part specific to the error function is novel to the best
of our knowledge.

For quadratic potentials, the most likely (self--consistent) measure
reads
\begin{align*}
\bar{\mu}(x) & =\lr{\delta(\dot{x}+x-\eta)}_{\eta},
\end{align*}
corresponding to the Gaussian process $\dot{x}=-x+\eta$, where $\eta$
is a zero--mean Gaussian process with self--consistent correlation
function
\begin{align*}
C_{\eta}(t_{1},t_{2}) & =2D\,\delta(t_{1}-t_{2})+g^{2}C_{\phi}(t_{1},t_{2})
\end{align*}
with $C_{\phi}(t_{1},t_{2})=\pint x\bar{\mu}(x)\,\phi(x(t_{1}))\phi(x(t_{2}))$.
Using the linearity of the dynamics of $x$, one obtains an ODE for
its stationary autocorrelation function $C_{x}(\tau)$,
\begin{align}
\ddot{C}_{x} & =C_{x}-g^{2}C_{\phi},\label{eq:sc_corr_ode}
\end{align}
with initial conditions $C_{x}(0)=\sigma_{x}^{2}$ and $\dot{C}_{x}(0)=-D$
\citep{Sompolinsky88_259,Schuecker18_041029}. Using Price's theorem,
\prettyref{eq:sc_corr_ode} can be cast into an equation of motion
$\ddot{C}_{x}=-\partial_{C_{x}}V(C_{x},\sigma_{x}^{2})$ in a potential
\begin{align}
V(C_{x},\sigma_{x}^{2}) & =-\frac{1}{2}C_{x}^{2}+g^{2}C_{\Phi}\label{eq:sc_corr_potential}
\end{align}
where $C_{\Phi}(t_{1},t_{2})=\pint x\bar{\mu}(x)\,\Phi(x(t_{1}))\Phi(x(t_{2}))$
and $\partial_{x}\Phi(x)=\phi(x)$.

Due to the implicit dependence of $C_{\Phi}$ on $C_{x}$ and $\sigma_{x}^{2}$,
this is not an initial value problem. To determine $\sigma_{x}^{2}$,
we use energy conservation $\frac{1}{2}\dot{C}_{x}^{2}+V(C_{x},\sigma_{x}^{2})=\mathrm{const}$.
We restrict ourselves to solutions where $C_{x}(\tau\to\infty)=0$
and $\dot{C}_{x}(\tau\to\infty)=0$. With this, energy conservation
evaluated at $\tau=0$ and $\tau\to\infty$ yields an equation for
$\sigma_{x}^{2}$:
\begin{align}
\frac{1}{2}D^{2}+V(\sigma_{x}^{2},\sigma_{x}^{2}) & =V(0,\sigma_{x}^{2}).\label{eq:sc_var_equation}
\end{align}
With $\sigma_{x}^{2}$ determined, \prettyref{eq:sc_corr_ode} becomes
an initial value problem that is straightforward to solve numerically.

Instead of solving \prettyref{eq:sc_var_equation} for given $D$
to get $\sigma_{x}^{2}$, we can use it to answer the inverse question:
Given $g$ and a desired activity level $\sigma_{x}^{2}$, how strong
does the external noise $D$ need to be? The answer directly follows
from \prettyref{eq:sc_var_equation}:
\begin{align}
D(\sigma_{x}^{2}) & =\sqrt{2(V(0,\sigma_{x}^{2})-V(\sigma_{x}^{2},\sigma_{x}^{2}))}.\label{eq:noise_given_var}
\end{align}
We use \prettyref{eq:noise_given_var} to uncover the multiple self--consistent
solutions; they correspond to a non--monotonicity of $D(\sigma_{x}^{2})$.
For arbitrary transfer functions, we solve the integrals for $C_{\Phi}$
numerically using an appropriate Gaussian quadrature.

\paragraph{Error Function}

For the transfer function
\begin{align*}
\phi(x) & =\erf(\sqrt{\pi}x/2),
\end{align*}
we can leverage an analytical expression for $C_{\phi}$ \citep[Appendix]{Williams98_1203}:
\begin{align}
C_{\phi}(\tau) & =\frac{2}{\pi}\,\arcsin\left(\frac{\pi C_{x}(\tau)}{2+\pi\sigma_{x}^{2}}\right).\label{eq:C_phi_erf}
\end{align}
For convenience, we introduce the scaled correlation function
\begin{align*}
y(\tau) & =\frac{\pi C_{x}(\tau)}{2+\pi\sigma_{x}^{2}},\qquad C_{x}(\tau)=\frac{2}{\pi}\frac{y(\tau)}{1-y_{0}}.
\end{align*}
Since $y$ depends linearly on $C_{x}$, we get from \prettyref{eq:sc_corr_ode}
an equation of motion for $y$,
\begin{align}
\ddot{y} & =y-g^{2}(1-y_{0})\arcsin\left(y\right),\label{eq:sc_y_ode}
\end{align}
with $y(0)\equiv y_{0}=\frac{\pi\sigma_{x}^{2}}{2+\pi\sigma_{x}^{2}}$
and $\dot{y}(0)=\frac{\pi}{2}(1-y_{0})D$ which again can be rewritten
as $\ddot{y}=-\partial_{y}V(y,y_{0})$. Using \prettyref{eq:C_phi_erf},
we get the explicit expression for the potential
\begin{align*}
V(y,y_{0}) & =-\frac{1}{2}y^{2}+g^{2}(1-y_{0})\left(\sqrt{1-y^{2}}+y\arcsin\left(y\right)-1\right).
\end{align*}
We chose the offset of the potential such that $V(0,y_{0})=0$ which
reduces \prettyref{eq:sc_var_equation} to
\begin{align}
\frac{\pi^{2}}{8}(1-y_{0})^{2}D^{2}+V(y_{0},y_{0}) & =0.\label{eq:sc_y0_eq}
\end{align}
We solve \prettyref{eq:sc_y0_eq} numerically using the Newton--Raphson
method implemented in SciPy \citep{Virtanen20_261} and \prettyref{eq:sc_y_ode}
using lsoda from the FORTRAN library odepack through the corresponding
SciPy interface.

From \prettyref{eq:sc_y_ode} we can determine the timescale of $y$
or equivalently $C_{x}$. Since $y(\tau\to\infty)\to0$, we linearize
\prettyref{eq:sc_y_ode} for $\tau\gg0$ to
\begin{align*}
\ddot{y} & =(1-g^{2}(1-y_{0}))y+O(y^{3}).
\end{align*}
From here, we can directly read off the timescale:
\begin{align}
\tau_{c} & =\frac{1}{\sqrt{1-g^{2}(1-y_{0})}}.\label{eq:y_timescale}
\end{align}
We use \prettyref{eq:y_timescale} to determine the timescale of the
prediction error (see below).

\subsubsection{Timescale of Prediction Error}

We here relate the timescale of the prediction error to the timescale
of the autocorrelation function $C_{x}(\tau)/C_{x}(0)\sim\exp(-\tau/\tau_{c})$.
The predicted variance in the continuous time limit is determined
by the corresponding limit of \prettyref{eq:pred_var},
\begin{align*}
\sigma_{\hat{x}}^{2} & =C_{x}(\hat{t},\hat{t})-\int_{0}^{\T}\int_{0}^{\T}C_{x}(\hat{t},u)C_{x}^{-1}(u,v)C_{x}(v,\hat{t})\,du\,dv,
\end{align*}
where $T$ denotes the training interval. Writing $\hat{t}=T+\tau$
and approximating $C_{x}(T+\tau,u)\approx C_{x}(T,u)e^{-\tau/\tau_{c}}$,
we get
\begin{align*}
\sigma_{\hat{x}}^{2} & \approx C_{x}(\hat{t},\hat{t})-e^{-2\tau/\tau_{c}}C_{x}(T,T),
\end{align*}
where we used $\int_{0}^{\T}C_{x}^{-1}(u,v)C_{x}(v,T)\,dv=\delta(u-T)$.
Using stationarity $C_{x}(u,v)=C_{x}(v-u)$, we arrive at
\begin{align*}
\sigma_{\hat{x}}^{2}/\sigma_{x}^{2} & \approx1-e^{-2\tau/\tau_{c}}
\end{align*}
where $C_{x}(0)=\sigma_{x}^{2}$. Thus, for large $\tau$, the timescale
of the prediction error is given by $\tau_{c}/2$.

\subsection{Fluctuations (Single Population)}

\subsubsection{Order Parameter Fluctuations}

Here, we derive an expression for the fluctuations of the variance
valid for slow dynamics $\tau_{c}\gg1$. According to \prettyref{eq:y_timescale},
this is valid for $g$ being of order $1$ - in practice, we choose
$g$ not too close to $1$, however, because of the periodic solutions
occurring in finite-size systems in this case \citep{Sompolinsky88_259}.
We start with the Legendre transform of the rate function of $C$,
\prettyref{eq:Rate_fct_C_contraction_principle}, which is the scaled
cumulant generating functional
\begin{align*}
W_{\infty}(k) & =-\C_{k}^{\T}\Ct_{k}+\C_{k}^{\T}k+\Omega_{0}(\C_{k},\Ct_{k}),\\
\Omega_{0}(C,\tilde{C}) & =\ln\,\pint x\pint{\xt}e^{S_{0}(x,\xt)+\frac{g^{2}}{2}\xt^{\T}C\xt+\phi^{\T}\tilde{C}\phi},\\
C_{k} & =\left.\partial_{\tilde{C}}\Omega_{0}(\C,\Ct)\right|_{C_{k},\tilde{C}_{k}},\\
\tilde{C}_{k} & =k+\left.\partial_{C}\Omega_{0}(\C,\Ct)\right|_{C_{k},\tilde{C}_{k}},
\end{align*}
where we redefined $\phi^{\T}k\phi\to k$ in the argument of $W_{\infty}$
to simplify the notation a bit. To determine the fluctuations, we
need to calculate the second derivative of the scaled cumulant generating
functional $W^{\prime\prime}(0)$.

We get immediately
\begin{align*}
W^{\prime}(k) & =C_{k}
\end{align*}
due to the saddle--point equations. The second derivative is thus
simply
\begin{align*}
W^{\prime\prime}(k) & =\frac{dC_{k}}{dk}.
\end{align*}
Using the saddle--point equations, we get
\begin{align*}
\left.\frac{dC_{k}}{dk}\right|_{C_{k},\tilde{C}_{k}} & =\left.\frac{dC_{k}}{dk}^{\T}\partial_{C}\partial_{\tilde{C}}\Omega_{0}(\C,\Ct)\right|_{C_{k},\tilde{C}_{k}}+\left.\frac{d\tilde{C}_{k}}{dk}^{\T}\partial_{\tilde{C}}\partial_{\tilde{C}}\Omega_{0}(\C,\Ct)\right|_{C_{k},\tilde{C}_{k}},\\
\left.\frac{d\tilde{C}_{k}}{dk}\right|_{C_{k},\tilde{C}_{k}} & =\delta+\left.\frac{dC_{k}}{dk}^{\T}\partial_{C}\partial_{C}\Omega_{0}(\C,\Ct)\right|_{C_{k},\tilde{C}_{k}}+\left.\frac{d\tilde{C}_{k}}{dk}^{\T}\partial_{\tilde{C}}\partial_{C}\Omega_{0}(\C,\Ct)\right|_{C_{k},\tilde{C}_{k}}.
\end{align*}
Evaluated at $k=0$ where $C_{k}=C_{0}$ and $\tilde{C}_{k}=0$, we
get
\begin{align*}
\left.\frac{dC_{k}}{dk}\right|_{C_{0},0} & =\frac{g^{2}}{2}\left.\frac{dC_{k}}{dk}\right|_{C_{0},0}^{\T}\ll\xt\xt,\phi\phi\rr_{0}+\left.\frac{d\tilde{C}_{k}}{dk}\right|_{C_{0},0}^{\T}\ll\phi\phi,\phi\phi\rr_{0},\\
\left.\frac{d\tilde{C}_{k}}{dk}\right|_{C_{0},0} & =\delta+\frac{g^{2}}{2}\left.\frac{d\tilde{C}_{k}}{dk}\right|_{C_{0},0}^{\T}\ll\phi\phi,\xt\xt\rr_{0},
\end{align*}
where we dropped $\ll\xt\xt,\xt\xt\rr_{0}=0$. The second equation
yields
\begin{align*}
\left.\frac{d\tilde{C}_{k}}{dk}\right|_{C_{0},0} & =A^{-1},\qquad A=\delta-\frac{g^{2}}{2}\ll\phi\phi,\xt\xt\rr_{0},
\end{align*}
inserting this in the first we get
\begin{align*}
\left.\frac{dC_{k}}{dk}\right|_{C_{0},0} & =A^{-1}\ll\phi\phi,\phi\phi\rr_{0}B^{-1},\qquad B=\delta-\frac{g^{2}}{2}\ll\xt\xt,\phi\phi\rr_{0}.
\end{align*}
We arrive at
\begin{align*}
W^{\prime\prime}(0) & =A^{-1}\ll\phi\phi,\phi\phi\rr_{0}B^{-1}.
\end{align*}
To avoid the complication of inverting the operators $A$ and $B$,
which depend on four times, we consider the implicit equation
\begin{align}
AW^{\prime\prime}(0)B & =\ll\phi\phi,\phi\phi\rr_{0}.\label{eq:fluct_A_B_equation}
\end{align}
Next, we simplify the operators $A$ and $B$.

First, we note that
\begin{align*}
\ll\phi(t_{1})\phi(t_{2}),\xt(s_{1})\xt(s_{2})\rr_{0} & \equiv\l\phi(t_{1})\phi(t_{2})\xt(s_{1})\xt(s_{2})\r_{0}-\l\phi(t_{1})\phi(t_{2})\r_{0}\l\xt(s_{1})\xt(s_{2})\r_{0}\\
 & =\l\phi(t_{1})\phi(t_{2})\xt(s_{1})\xt(s_{2})\r_{0}
\end{align*}
because $\l\xt\xt\r_{0}=0$. Furthermore, because $\l\cdot\r_{0}$
is a Gaussian measure, we have 
\begin{align*}
\l\phi(t_{1})\phi(t_{2})\xt(s_{1})\xt(s_{2})\r_{0} & =\l\phi^{\prime\prime}(t_{1})\phi(t_{2})\r_{0}\l x(t_{1})\xt(s_{1})\r_{0}\l x(t_{1})\xt(s_{2})\r_{0}\\
 & +\l\phi^{\prime}(t_{1})\phi^{\prime}(t_{2})\r_{0}\l x(t_{1})\xt(s_{1})\r_{0}\l x(t_{2})\xt(s_{2})\r_{0}\\
 & +\l\phi^{\prime}(t_{1})\phi^{\prime}(t_{2})\r_{0}\l x(t_{2})\xt(s_{1})\r_{0}\l x(t_{1})\xt(s_{2})\r_{0}\\
 & +\l\phi(t_{1})\phi^{\prime\prime}(t_{2})\r_{0}\l x(t_{2})\xt(s_{1})\r_{0}\l x(t_{2})\xt(s_{2})\r_{0},
\end{align*}
which can be derived by expanding $\phi(x(t_{1}))$ and $\phi(x(t_{2}))$
as a Taylor series and applying Wick's theorem. The expectation $\l x(t_{1})\xt(t_{2})\r_{0}$
is the response at $t_{1}$ to an infinitesimal perturbation at $t_{2}$.

For quadratic potentials $U(x)=\frac{1}{2}x^{2}$, the linear response
function is $\l x(t_{1})\xt(t_{2})\r_{0}=-H(t_{1}-t_{2})e^{-(t_{1}-t_{2})}$.
In particular, its timescale is given by the timescale of the single
unit dynamics, i.e.~unity in the dimensionless units. In contrast,
the timescale of the other expectations $\l\cdot\r_{0}$ is determined
by the timescale of $C_{x}$, i.e.~$\tau_{c}$. For $\tau_{c}\gg1$,
$W^{\prime\prime}(0)$ hardly changes on the timescale of $\l x\xt\r_{0}$,
thus we can approximate $\l x(t_{1})\xt(t_{2})\r_{0}\approx-\delta(t_{2}-t_{1})$.
Because we are only interested in the fluctuations of the variance,
we furthermore evaluate \prettyref{eq:fluct_A_B_equation} at equal
times and consider the stationary case. This turns the contributions
to $A$ and $B$ dependent on $\phi$ and its derivatives into constants
and, most notably, renders \prettyref{eq:fluct_A_B_equation} independent
of time. We therefore suppress the time argument again to arrive at
\begin{align*}
\langle\Delta q^{2}\rangle & =\frac{\ll\phi\phi,\phi\phi\rr_{0}}{N\left(1-g^{2}(\l\phi^{\prime\prime}\phi\r_{0}+\l\phi^{\prime}\phi^{\prime}\r_{0})\right)^{2}}
\end{align*}
as stated in the main text. The factor $1/N$ is due to the definition
of the scaled cumulant generating functional, $W_{\infty}(k)=\lim_{N\to\infty}\frac{1}{N}\ln\lr{\lr{e^{NC^{\T}k}}_{\vect x|\matr J}}_{\matr J}$,
where the factor $N$ in the exponent generates a factor $N$ with
each derivative of $W_{\infty}$. Conversely, the derivatives of $W_{\infty}$
yields the n--th cumulant scaled with $1/N^{n-1}$. Lastly, we used
$\ll\phi\phi,\phi\phi\rr_{0}\equiv\l\phi\phi\phi\phi\r_{0}-\l\phi\phi\r_{0}\l\phi\phi\r_{0}=\l(\phi\phi-\l\phi\phi\r_{0})^{2}\r_{0}$
in the main text.

\begin{figure}
\includegraphics{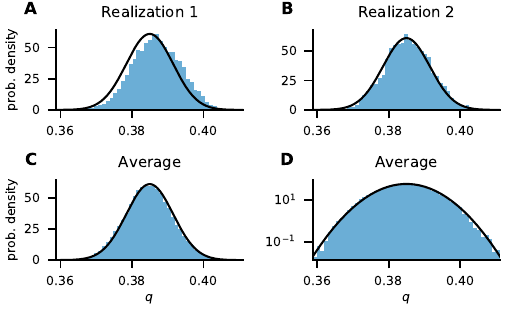}

\caption{Order parameter fluctuations. \textbf{A},\textbf{B} Temporal order
parameter statistics for a single realization of random connectivity
each. \textbf{C},\textbf{D} Temporal order parameter statistics across
ten simulations with linear and logarithmic y-axis (panel \textbf{D}
is identical to Fig.~2\textbf{A} in the main text). \label{fig:order_param_fluct}}
\end{figure}

In the main text, we show the fluctuations of the order parameter
across time and realizations of the connectivity in Fig.~2\textbf{A}.
To supplement this, we show the order parameter fluctuations in \prettyref{fig:order_param_fluct}
for two realizations of the connectivity (\prettyref{fig:order_param_fluct}\textbf{A},\textbf{B})
and averaged across ten realizations of the connectivity (\prettyref{fig:order_param_fluct}\textbf{C},\textbf{D}).
Using a logarithmic y-axis reveals that also the tails are Gaussian.

\subsubsection{Coexisting Mean--Field Solutions}

Here, we determine a regime where two mean--field solutions coexist.
We restrict ourselves to quadratic potentials $U(x)=\frac{1}{2}x^{2}$
and start from \prettyref{eq:noise_given_var},
\begin{align*}
D(\sigma_{x}^{2}) & =\sqrt{2(V(0,\sigma_{x}^{2})-V(\sigma_{x}^{2},\sigma_{x}^{2}))},
\end{align*}
which determines the necessary external noise to reach a given activity
level $\sigma_{x}^{2}$. Non--monotonicities of $D(\sigma_{x}^{2})$
give rise to multiple solutions since they indicate a case where the
same external noise can lead to different activity levels.

We focus on the linearly stable case $g<1$ with antisymmetric $\phi(x)$
and $\phi^{\prime}(0)=1$. For small $\sigma_{x}^{2}$, we approximate
$\Phi(x)=\frac{1}{2}x^{2}+\frac{\alpha}{24}x^{4}+O(x^{6})$. Using
Wick's theorem and \prettyref{eq:sc_corr_potential}, we get
\[
2(V(0,\sigma_{x}^{2})-V(\sigma_{x}^{2},\sigma_{x}^{2}))=(1-g^{2})\sigma_{x}^{4}-\alpha g^{2}\sigma_{x}^{6}+O(\sigma_{x}^{8}).
\]
For $g<1$, the leading order term grows monotonically with $\sigma_{x}$.
To introduce a non--monotonicity, the next term has to shrink which
implies $\alpha>0$. This excludes sigmoidal functions like $\phi(x)=\erf(\sqrt{\pi}x/2)$
or $\phi(x)=\tanh(x)$. Thus, we consider non--sigmoidal functions
with $\alpha>1$ that we keep bounded between $-1$ and $1$ by clipping
them to the interval $[-1,1]$.

In the noiseless case $D=0$, the silent fixed point $\sigma_{x}^{2}=0$
is one of the two solutions. Using the stability criterion $g^{2}\langle\phi^{\prime}(\sigma_{x})^{2}\rangle<1$
from \citep{Kadmon15_041030}, we get for the transfer function $\phi(x)=\clip(\tan(x),-1,1)$
\[
g^{2}\langle\phi^{\prime}(\sigma_{x}z)^{2}\rangle=g^{2}\int_{-\pi/4}^{\pi/4}dz\,\mathcal{N}(z\,|\,0,\sigma_{x}^{2})\,\cos^{-4}(z)\overset{\sigma_{x}\to0}{\to}g^{2},
\]
hence the silent fixed point is stable for $g<1$.

\subsection{Rate Function (Multiple Populations)}

\subsubsection{Scaled Cumulant Generating Functional}

Here, we derive the scaled cumulant generating functional and the
saddle-point equations for networks with multiple populations. The
steps are similar to the single population case, hence we keep the
presentation brief. Throughout, we use greek indices for the populations
and latin indices for individual neurons within a given population:
$x_{i}^{\alpha}$ denotes the trajectory of neuron $i$ of population
$\alpha$, $\vect x^{\alpha}$ the trajectories of all neurons in
population $\alpha$, and $\vect x$ the trajectories of all neurons.
The same convention applies to the connectivity: $J_{ij}^{\alpha\beta}$
governs the connection from neuron $j$ in population $\beta$ to
neuron $i$ in population $\alpha$, $\matr J^{\alpha\beta}$ the
connections from all neurons in population $\beta$ to all neurons
in population $\alpha$, and $\matr J$ all connections. Furthermore,
we denote the size of an individual population by $N_{\alpha}$ and
set $N=\sum_{\alpha}N_{\alpha}$.

The expectation $\lr{\cdot}_{\vect x|\matr J}$ of some arbitrary
functional $G(\vect x)$ can again be written as
\begin{align*}
\lr{\lr{G(\vect x)}_{\vect x|\matr J,\vect{\xi}}}_{\vect{\xi}} & =\prod_{\alpha}\pint{\vect x^{\alpha}}P(\vect x\,|\,\matr J)\,G(\vect x),
\end{align*}
where we introduced
\begin{align*}
P(\vect x\,|\,\matr J) & =\prod_{\alpha}\lr{\delta(\tau_{\alpha}\dot{\vect x}^{\alpha}+U_{\alpha}^{\prime}(\vect x^{\alpha})+\sum_{\beta}\matr J^{\alpha\beta}\phi(\vect x^{\beta})+\vect{\xi}^{\alpha})}_{\vect{\xi}^{\alpha}}\\
 & =\prod_{\alpha}\pint{\vect{\tilde{x}}^{\alpha}}\,e^{\sum_{\alpha}S_{0}^{\alpha}(\vect x^{\alpha},\vect{\tilde{x}}^{\alpha})-\sum_{\alpha,\beta}\vect{\tilde{x}}^{\alpha\T}\matr J^{\alpha\beta}\phi(\vect x^{\beta})}.
\end{align*}
The action $S_{0}^{\alpha}$ now depends on the population,
\begin{align*}
S_{0}^{\alpha}(\vect y,\vect{\tilde{y}}) & =\vect{\tilde{y}}^{\T}(\tau_{\alpha}\dot{\vect y}+U_{\alpha}^{\prime}(\vect y))+D_{\alpha}\vect{\tilde{y}}^{\T}\vect{\tilde{y}}.
\end{align*}
The average over realizations of the connectivity $\matr J^{\alpha\beta}\stackrel{\text{i.i.d.}}{\sim}\mathcal{N}(0,N_{\beta}^{-1}g_{\alpha\beta}^{2})$
only affects the term $-\sum_{\alpha,\beta}\vect{\tilde{x}}^{\alpha\T}\matr J^{\alpha\beta}\phi(\vect x^{\beta})$.
Due to the independence of the entries of $\matr J$, the average
factorizes into
\begin{align*}
\l e^{-\sum_{\alpha,\beta}\vect{\tilde{x}}^{\alpha\T}\matr J^{\alpha\beta}\phi(\vect x^{\beta})}\r_{\matr J} & =\prod_{\alpha,i}\prod_{\beta,j}\l e^{-\tilde{x}_{i}^{\alpha\T}J_{ij}^{\alpha\beta}\phi(x_{j}^{\beta})}\r_{J_{ij}^{\alpha\beta}}=\prod_{\alpha,i}e^{\frac{1}{2}\tilde{x}_{i}^{\alpha\T}\left(\sum_{\beta,j}\frac{g_{\alpha\beta}^{2}}{N_{\beta}}\phi(x_{j}^{\beta})\phi(x_{j}^{\beta})^{\T}\right)\tilde{x}_{i}^{\alpha}}.
\end{align*}
Next, we introduce the population--averaged auxiliary fields
\begin{align*}
C^{\alpha}(u,v) & =\frac{1}{N_{\alpha}}\sum_{i=1}^{N_{\alpha}}\phi(x_{i}^{\alpha}(u))\phi(x_{i}^{\alpha}(v))
\end{align*}
via Hubbard--Stratonovich transformations:
\begin{align*}
\l e^{-\sum_{\alpha,\beta}\vect{\tilde{x}}^{\alpha\T}\matr J^{\alpha\beta}\phi(\vect x^{\beta})}\r_{\matr J} & =\prod_{\alpha}\pint{C^{\alpha}}\pint{\tilde{C}^{\alpha}}e^{-\sum_{\alpha}N_{\alpha}\,C^{\alpha\T}\tilde{C}^{\alpha}+\sum_{\alpha}\phi(\vect x^{\alpha})^{\T}\tilde{C}^{\alpha}\phi(\vect x^{\alpha})+\frac{1}{2}\sum_{\alpha}\vect{\tilde{x}}^{\alpha\T}(\sum_{\beta}g_{\alpha\beta}^{2}C^{\beta})\vect{\tilde{x}}^{\alpha}}.
\end{align*}
As in the single-population case, the average over the connectivity
and the subsequent Hubbard--Stratonovich transformation decouple
the dynamics across units; afterwards, the units are only coupled
through the global fields $C^{\alpha}$ and $\tilde{C}^{\alpha}$.

Now, we consider the empirical densities of the populations,
\begin{align}
\mu^{\alpha}(y) & =\frac{1}{N_{\alpha}}\sum_{i=1}^{N_{\alpha}}\delta(x_{i}^{\alpha}-y).\label{eq:def_mu_multipop}
\end{align}
 The corresponding scaled cumulant generating functional is
\begin{align}
W_{N}(\{\ell^{\circ}\}) & =\frac{1}{N}\ln\lr{\lr{e^{\sum_{\alpha}\sum_{i=1}^{N_{\alpha}}\k^{\alpha}(x_{i})}}_{\vect x|\matr J}}_{\matr J},\label{eq:def_W_multipop}
\end{align}
where we introduced one functional $\ell^{\alpha}$ for each $\mu^{\alpha}$
and the collection of all $\ell^{\alpha}$, $\{\ell^{\circ}\}$. Using
the above results and the abbreviation $\phi(x)\equiv\phi$, it can
be written as
\begin{align*}
W_{N}(\{\ell^{\circ}\})=\frac{1}{N}\ln\,\prod_{\alpha}\pint{C^{\alpha}} & \pint{\tilde{C}^{\alpha}}e^{-\sum_{\alpha}N_{\alpha}\,C^{\alpha\T}\tilde{C}^{\alpha}+\sum_{\alpha}N_{\alpha}\,\Omega_{\ell^{\alpha}}^{\alpha}(\{C^{\circ}\},\tilde{C}^{\alpha})},
\end{align*}
where we introduced
\begin{align*}
\Omega_{\ell}^{\alpha}(\{C^{\circ}\},\tilde{C})=\ln\,\pint x\pint{\xt} & e^{S_{0}^{\alpha}(x,\xt)+\frac{1}{2}\xt^{\T}(\sum_{\beta}g_{\alpha\beta}^{2}C^{\beta})\xt+\phi^{\T}\tilde{C}\phi+\k(x)}.
\end{align*}
Again, the $N_{\alpha}$ in front of the single--particle cumulant
generating functionals $\Omega_{\ell}^{\alpha}$ results from the
factorization of the $N_{\alpha}$ integrals over $x_{i}^{\alpha}$
and $\tilde{x}_{i}^{\alpha}$ each; thus it is a hallmark of the decoupled
dynamics. Note that $W_{N}(\{\ell^{\circ}\})$ is still coupled across
populations, because each $\Omega_{\ell}^{\alpha}$ depends on the
set of all auxiliary fields, $\{C^{\circ}\}$.

Next, we approximate the $C^{\alpha}$ and $\tilde{C}^{\alpha}$ integrals
in a saddle--point approximation which yields
\begin{align}
W_{\infty}(\{\ell^{\circ}\}) & =-\sum_{\alpha}\gamma_{\alpha}C_{\{\ell^{\circ}\}}^{\alpha\T}\tilde{C}_{\{\ell^{\circ}\}}^{\alpha}+\sum_{\alpha}\gamma_{\alpha}\,\Omega_{\ell^{\alpha}}^{\alpha}(\{C_{\{\ell^{\circ}\}}^{\circ}\},\tilde{C}_{\{\ell^{\circ}\}}^{\alpha}),\label{eq:W_inf_multipop}
\end{align}
where $\gamma_{\alpha}=N_{\alpha}/N$. $C_{\{\ell^{\circ}\}}^{\alpha}$
and $\tilde{C}_{\{\ell^{\circ}\}}^{\alpha}$ are determined by the
saddle--point equations
\begin{align}
C_{\{\ell^{\circ}\}}^{\alpha} & =\evalat{\partial_{\Ct}\Omega_{\ell^{\alpha}}^{\alpha}(\{C^{\circ}\},\Ct)}_{\{C_{\{\ell^{\circ}\}}^{\circ}\},\tilde{C}_{\{\ell^{\circ}\}}^{\alpha}},\label{eq:C_spe_multipop}\\
\gamma_{\alpha}\tilde{C}_{\{\ell^{\circ}\}}^{\alpha} & =\sum_{\beta}\gamma_{\beta}\evalat{\partial_{\C^{\alpha}}\Omega_{\ell^{\beta}}^{\beta}(\{C^{\circ}\},\tilde{C})}_{\{C_{\{\ell^{\circ}\}}^{\circ}\},\tilde{C}_{\{\ell^{\circ}\}}^{\beta}}.\label{eq:Ct_spe_multipop}
\end{align}
Here, the asymmetry in the saddle-point equations reflects the fact
that $\Omega_{\ell}^{\alpha}$ depends on a single $\tilde{C}$ but
on all $\{C^{\circ}\}$.

\subsubsection{Rate Function}

Here, we derive the rate function from the scaled cumulant generating
functional for the multi-population case. We obtain the rate function
via the Legendre transformation
\begin{align}
H(\{\mu^{\circ}\}) & =\sum_{\alpha}\gamma_{\alpha}\pint x\mu^{\alpha}(x)\k_{\{\mu^{\circ}\}}^{\alpha}(x)-W_{\infty}(\{\k_{\{\mu^{\circ}\}}^{\circ}\})\label{eq:def_H_Legendre_multipop}
\end{align}
with $\k_{\{\mu^{\circ}\}}^{\alpha}$ implicitly defined by
\begin{align}
\gamma_{\alpha}\mu^{\alpha} & =\left.\partial_{\ell^{\alpha}}W_{\infty}(\{\ell^{\circ}\})\right|_{\{\k_{\{\mu^{\circ}\}}^{\circ}\}}.\label{eq:implicit_eq_ell_mu_multipop}
\end{align}
Due to the saddle--point equations, \prettyref{eq:C_spe_multipop}
and \prettyref{eq:Ct_spe_multipop}, the derivative of the cumulant
generating functional in \prettyref{eq:implicit_eq_ell_mu_multipop}
simplifies to 
\begin{align*}
\left.\partial_{\ell^{\alpha}}W_{\infty}(\{\ell^{\circ}\})\right|_{\{\k_{\{\mu^{\circ}\}}^{\circ}\}} & =\evalat{\gamma_{\alpha}\,\partial_{\ell^{\alpha}}\Omega_{\ell^{\alpha}}^{\alpha}(\{C_{\{\ell^{\circ}\}}^{\circ}\},\tilde{C}_{\{\ell^{\circ}\}}^{\alpha})}_{\{\k_{\{\mu^{\circ}\}}^{\circ}\}},
\end{align*}
where the derivative only acts on the $\ell^{\alpha}$ that is explicit
in $\Omega_{\ell^{\alpha}}^{\alpha}$ and not on the implicit dependencies
through $\{C_{\{\ell^{\circ}\}}^{\circ}\}$, $\tilde{C}_{\{\ell^{\circ}\}}^{\alpha}$.
Thus, \prettyref{eq:implicit_eq_ell_mu_multipop} yields
\begin{align}
\mu^{\alpha}(x) & =\left.\frac{\lr{\delta(\tau_{\alpha}\dot{x}+U_{\alpha}^{\prime}(x)-\eta_{\alpha})}_{\eta_{\alpha}}e^{\phi^{\T}\tilde{C}_{\{\ell^{\circ}\}}^{\alpha}\phi+\k^{\alpha}(x)}}{\pint x\lr{\delta(\tau_{\alpha}\dot{x}+U_{\alpha}^{\prime}(x)-\eta_{\alpha})}_{\eta_{\alpha}}e^{\phi^{\T}\tilde{C}_{\{\ell^{\circ}\}}^{\alpha}\phi+\k^{\alpha}(x)}}\right|_{\{\k_{\{\mu^{\circ}\}}^{\circ}\}},\label{eq:mu_alpha}
\end{align}
where we used 
\begin{align*}
\pint{\xt}e^{\tilde{x}^{\T}(\tau\dot{x}+U^{\prime}(x))+\frac{1}{2}\xt^{\T}C_{\eta}\xt} & =\lr{\delta(\tau\dot{x}+U^{\prime}(x)-\eta)}_{\eta}
\end{align*}
to introduce the zero-mean Gaussian process $\eta_{\alpha}$ with
correlation function
\begin{align*}
C_{\eta}^{\alpha} & (u,v)=2D_{\alpha}\delta(u-v)+\sum_{\beta}g_{\alpha\beta}^{2}C_{\{\k_{\{\mu^{\circ}\}}^{\circ}\}}^{\beta}(u,v).
\end{align*}
Taking the logarithm of \prettyref{eq:mu_alpha} and using the definition
of $\Omega_{\ell}^{\alpha}$ leads to
\begin{align*}
\k_{\{\mu^{\circ}\}}^{\alpha}(x)= & \ln\frac{\mu^{\alpha}(x)}{\lr{\delta(\tau_{\alpha}\dot{x}+U_{\alpha}^{\prime}(x)-\eta_{\alpha})}_{\eta_{\alpha}}}-\phi^{\T}\tilde{C}_{\{\k_{\{\mu^{\circ}\}}^{\circ}\}}^{\alpha}\phi+\left.\Omega_{\ell^{\alpha}}^{\alpha}(\{C_{\{\ell^{\circ}\}}^{\circ}\},\tilde{C}_{\{\ell^{\circ}\}}^{\alpha})\right|_{\{\k_{\{\mu^{\circ}\}}^{\circ}\}}.
\end{align*}
Inserting $\k_{\{\mu^{\circ}\}}^{\alpha}(x)$ into the Legendre transformation
\eqref{eq:def_H_Legendre_multipop} and using \prettyref{eq:W_inf_multipop}
as $\sum_{\alpha}\gamma_{\alpha}\,\Omega_{\ell^{\alpha}}^{\alpha}(\{C_{\{\ell^{\circ}\}}^{\circ}\},\tilde{C}_{\{\ell^{\circ}\}}^{\alpha})-W_{\infty}(\{\ell^{\circ}\})=\sum_{\alpha}\gamma_{\alpha}C_{\{\ell^{\circ}\}}^{\alpha\T}\tilde{C}_{\{\ell^{\circ}\}}^{\alpha}$
yields
\begin{align*}
H(\{\mu^{\circ}\})= & \sum_{\alpha}\gamma_{\alpha}\pint x\mu^{\alpha}(x)\ln\frac{\mu^{\alpha}(x)}{\lr{\delta(\tau_{\alpha}\dot{x}+U_{\alpha}^{\prime}(x)-\eta_{\alpha})}_{\eta_{\alpha}}}-\sum_{\alpha}\gamma_{\alpha}C_{\mu^{\alpha}}^{\T}\tilde{C}_{\{\k_{\{\mu^{\circ}\}}^{\circ}\}}^{\alpha}+\sum_{\alpha}\gamma_{\alpha}C_{\{\k_{\{\mu^{\circ}\}}^{\circ}\}}^{\alpha\T}\tilde{C}_{\{\k_{\{\mu^{\circ}\}}^{\circ}\}}^{\alpha},
\end{align*}
where
\begin{align*}
C_{\mu^{\alpha}}(u,v) & =\pint x\mu^{\alpha}(x)\phi(x(u))\phi(x(v)).
\end{align*}
Identifying $\mu^{\alpha}(x)$ in the saddle--point equation \eqref{eq:C_spe_multipop}
yields
\begin{align*}
C_{\{\k_{\{\mu^{\circ}\}}^{\circ}\}}^{\alpha}(u,v) & =\pint x\mu^{\alpha}(x)\phi(x(u))\phi(x(v))
\end{align*}
and thus $C_{\{\k_{\{\mu^{\circ}\}}^{\circ}\}}^{\alpha}=C_{\mu^{\alpha}}$.
Accordingly, the last two terms in the Legendre transformation cancel
and we arrive at
\begin{equation}
H(\{\mu^{\circ}\})=\sum_{\alpha}\gamma_{\alpha}\pint x\mu^{\alpha}(x)\ln\frac{\mu^{\alpha}(x)}{\lr{\delta(\tau_{\alpha}\dot{x}+U_{\alpha}^{\prime}(x)-\eta_{\alpha})}_{\eta_{\alpha}}},\label{eq:rate_fct_multipop_appendix}
\end{equation}
where $\eta_{\alpha}$ is a zero-mean Gaussian process with correlation
function 
\begin{align}
C_{\eta}^{\alpha}(u,v) & =2D_{\alpha}\delta(u-v)+\sum_{\beta}g_{\alpha\beta}^{2}\pint x\mu^{\beta}(x)\phi(x(u))\phi(x(v)).\label{eq:C_eta_multipop_appendix}
\end{align}
Note that although \prettyref{eq:rate_fct_multipop_appendix} is a
sum over the populations, the individual terms are still coupled through
\prettyref{eq:C_eta_multipop_appendix}.

The derivation can be generalized further to population-specific transfer
functions $\phi_{\alpha}(x_{i}^{\alpha})$. Since this would make
the notation more complicated without any conceptual changes, we just
state the result: The rate function is still given by \prettyref{eq:rate_fct_multipop_appendix}
but the correlation function of $\eta_{\alpha}$ becomes
\begin{align*}
C_{\eta}^{\alpha}(u,v) & =2D_{\alpha}\delta(u-v)+\sum_{\beta}g_{\alpha\beta}^{2}\pint x\mu^{\beta}(x)\phi_{\beta}(x(u))\phi_{\beta}(x(v)).
\end{align*}
In the main text, we state only the slightly less general result for
$\phi_{\alpha}\equiv\phi$.

\subsection{Inference (Multiple Populations)}

\subsubsection{Log--Likelihood Derivative}

Here, we calculate the derivatives of the log--likelihood with respect
to the parameters $g_{\alpha\beta}$ and $D_{\alpha}$ for the multi-population
case. We denote the matrix with elements $g_{\alpha\beta}$ by $\bm{g}$
and the vector with elements $D_{\alpha}$ by $\bm{D}$ and proceed
similar to the single population case.

In terms of the rate function, \prettyref{eq:rate_fct_multipop_appendix},
we have
\begin{align*}
\partial_{a_{\alpha}}\ln P(\{\mu^{\circ}\}\,|\,\bm{g},\bm{D}) & \simeq-N\partial_{a_{\alpha}}H(\{\mu^{\circ}\}\,|\,\bm{g},\bm{D})
\end{align*}
where $a_{\alpha}$ denotes either $g_{\alpha\beta}$ and $D_{\alpha}$.
The parameters $a_{\alpha}$ appear only in the cross entropy of population
$\alpha$
\begin{align*}
\partial_{a_{\alpha}}H(\{\mu^{\circ}\})=-\gamma_{\alpha} & \pint x\mu^{\alpha}(x)\,\partial_{a_{\alpha}}\ln\lr{\delta(\tau_{\alpha}\dot{x}+U_{\alpha}^{\prime}(x)-\eta_{\alpha})}_{\eta_{\alpha}}
\end{align*}
through the correlation function $C_{\eta}^{\alpha}(u,v)=2D_{\alpha}\delta(u-v)+\sum_{\beta}g_{\alpha\beta}^{2}\pint x\mu^{\beta}(x)\phi(x(u))\phi(x(v))$.
In the calculation for the log-likelihood derivative for the single
population, we showed that
\begin{align*}
\lr{\delta(\tau\dot{x}+U^{\prime}(x)-\eta)}_{\eta} & =\frac{e^{-\frac{1}{2}(\tau\dot{x}+U^{\prime}(x))^{\T}C_{\eta}^{-1}(\tau\dot{x}+U^{\prime}(x))}}{\sqrt{\det(2\pi C_{\eta})}}.
\end{align*}
With this, we can take the derivatives of $\ln\lr{\delta(\tau_{\alpha}\dot{x}+U_{\alpha}^{\prime}(x)-\eta_{\alpha})}_{\eta_{\alpha}}$
and get
\begin{align*}
\partial_{a_{\alpha}}\ln\lr{\delta(\tau_{\alpha}\dot{x}+U_{\alpha}^{\prime}(x)-\eta_{\alpha})}_{\eta_{\alpha}} & =-\frac{1}{2}(\tau_{\alpha}\dot{x}+U_{\alpha}^{\prime}(x))^{\T}\frac{\partial(C_{\eta}^{\alpha})^{-1}}{\partial a_{\alpha}}(\tau_{\alpha}\dot{x}+U_{\alpha}^{\prime}(x))-\frac{1}{2}\partial_{a_{\alpha}}\tr\ln C_{\eta}^{\alpha},
\end{align*}
where we used $\ln\det C=\tr\ln C$. With this, we arrive at
\begin{align*}
\partial_{a_{\alpha}}H(\{\mu^{\circ}\}) & =\frac{\gamma_{\alpha}}{2}\tr\left(C_{0}^{\alpha}\frac{\partial(C_{\eta}^{\alpha})^{-1}}{\partial a_{\alpha}}\right)+\frac{\gamma_{\alpha}}{2}\tr\left(\frac{\partial C_{\eta}^{\alpha}}{\partial a_{\alpha}}(C_{\eta}^{\alpha})^{-1}\right),
\end{align*}
where the integral over the empirical measure $\mu^{\alpha}$ gave
rise to $C_{0}^{\alpha}=\pint x\mu^{\alpha}(x)(\tau_{\alpha}\dot{x}+U_{\alpha}^{\prime}(x))(\tau_{\alpha}\dot{x}+U_{\alpha}^{\prime}(x))$
and we used $\partial_{a}\ln C=\frac{\partial C}{\partial a}C^{-1}$.
Finally, using $\frac{\partial C}{\partial a}C^{-1}=CC^{-1}\frac{\partial C}{\partial a}C^{-1}=-C\frac{\partial C^{-1}}{\partial a}$,
we get
\begin{align}
\partial_{a}\ln P(\{\mu^{\alpha}\}\,|\,\bm{g},\bm{D}) & \simeq-\frac{N_{\alpha}}{2}\tr\left((C_{0}^{\alpha}-C_{\eta}^{\alpha})\frac{\partial(C_{\eta}^{\alpha})^{-1}}{\partial a_{\alpha}}\right).\label{eq:loglikelihood_drv_multipop_appendix}
\end{align}
The derivative vanishes for $C_{0}^{\alpha}=C_{\eta}^{\alpha}$.

Assuming stationarity, a Fourier transformation of $C_{0}^{\alpha}=C_{\eta}^{\alpha}$
leads to
\begin{align}
\mathcal{S}_{\tau_{\alpha}\dot{x}+U_{\alpha}^{\prime}(x)}^{\alpha}(f) & =2D_{\alpha}+\sum_{\beta}g_{\alpha\beta}^{2}\mathcal{S}_{\phi(x)}^{\beta}(f)\label{eq:inference_condition_multipop_appendix}
\end{align}
as stated in the main text.

\subsubsection{Degeneracy of Inference Equation}

\begin{figure}
\includegraphics{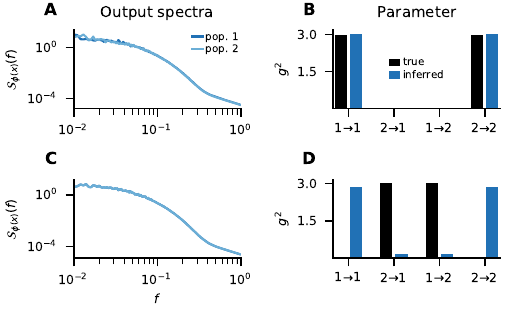}

\caption{Maximum likelihood parameter estimation for two populations with equal
time constants $\tau_{1}=\tau_{2}=1$ and equal quadratic row sums
$\sum_{\beta}g_{\alpha\beta}^{2}=3\quad\forall\alpha$. \textbf{A}
Output power spectra $\mathcal{S}_{\phi(x)}^{\alpha}(f)$ of two unconnected
populations $g_{12}^{2}=g_{21}^{2}=0$ with $g_{11}^{2}=g_{22}^{2}=3$.
\textbf{B} Estimated (blue) and true (black) parameters corresponding
to A. \textbf{C} Output power spectra of two connected populations
with $g_{11}^{2}=g_{22}^{2}=0$, $g_{12}^{2}=g_{21}^{2}=3$. \textbf{D}
Estimated (blue) and true (black) parameters corresponding to C. Further
parameters as in Fig.~3 in the main text. \label{fig:inference_degeneracy}}
\end{figure}

Here, we show that parameter inference using \prettyref{eq:inference_condition_multipop_appendix}
can be degenerate because different models are equally plausible.

If the empirical estimates of the output spectra agree, $\mathcal{S}_{\phi(x)}^{\alpha}(f)=\mathcal{S}_{\phi(x)}^{\beta}(f)\equiv\mathcal{S}_{\phi(x)}^{\beta}(f)$,
\prettyref{eq:inference_condition_multipop_appendix} reduces to
\begin{align*}
\mathcal{S}_{\tau_{\alpha}\dot{x}+U_{\alpha}^{\prime}(x)}^{\alpha}(f) & =2D_{\alpha}+\mathcal{S}_{\phi(x)}(f)\sum_{\beta}g_{\alpha\beta}^{2}.
\end{align*}
Clearly, this leads to a degenerate space of solutions with $\sum_{\beta}g_{\alpha\beta}^{2}=\mathrm{const.}$

For example, we consider the case with $\tau_{1}=\tau_{2}=1$ and
$\sum_{\beta}g_{\alpha\beta}^{2}=3$ in \prettyref{fig:inference_degeneracy}.
The most likely set of empirical measures for these parameters is
$\bar{\mu}^{\alpha}=\bar{\mu}^{\beta}$, hence the most likely empirical
output spectra agree. Indeed, the empirical output spectra of the
two populations agree almost perfectly for a given realization of
the connectivity (\prettyref{fig:inference_degeneracy}\textbf{A},\textbf{C}),
thereby rendering the inference degenerate. Accordingly, for two populations
without self-connections, $g_{11}^{2}=g_{22}^{2}=0$, $g_{12}^{2}=g_{21}^{2}=3$,
the parameter inference infers the opposite of two almost unconnected
populations (\prettyref{fig:inference_degeneracy}\textbf{C},\textbf{D}).
Curiously, the inferred parameters agree perfectly with the true parameters
if the populations are unconnected (\prettyref{fig:inference_degeneracy}\textbf{A},\textbf{B}).
This is a finite-size effect: For unconnected networks, the estimates
of the output spectra are independent, which leads to different finite-size
fluctuations (compare \prettyref{fig:inference_degeneracy}\textbf{A}
and \prettyref{fig:inference_degeneracy}\textbf{C}) such that the
inference is not degenerate anymore.